\documentstyle[12pt,epsfig]{article}
\begin{document}
\setlength{\parskip}{0.45cm}
\setlength{\baselineskip}{0.75cm}
%
%
%
\begin{titlepage}
\setlength{\parskip}{0.25cm}
\setlength{\baselineskip}{0.25cm}
\begin{flushright}
DO-TH 99/03\\  
\vspace{0.2cm}
hep--ph/9903337\\
\vspace{0.2cm}
February 1999
\end{flushright}
\vspace{1.0cm}
\begin{center}
\LARGE
{\bf Radiatively Generated Parton Distributions\\
\vspace{0.1cm}
 of Real and Virtual Photons}
\vspace{1.5cm}

\large
M. Gl\"uck, E.\ Reya and I.\ Schienbein\\
\vspace{1.0cm}

\normalsize
{\it Institut f\"{u}r Physik, Universit\"{a}t Dortmund}\\ 
{\it D-44221 Dortmund, Germany} \\

\vspace{1.5cm}
\end{center}
\begin{abstract}
The parton content of real ($P^2=0$) and virtual ($P^2\neq 0$) transverse
photons $\gamma(P^2)$ is expressed in terms of perturbative pointlike and
nonperturbative hadronic (VMD) components, employing recently updated
parton distributions of pions and protons.  The resulting parameter--free
and perturbatively stable LO and NLO parton densities $f^{\gamma(P^2)}
(x,Q^2)$ are smooth in $P^2$ and apply to all $P^2\geq0$ when-\- ever
$\gamma(P^2)$ is probed at scales $Q^2\gg P^2$ where transverse photons
also dominate physically relevant cross sections.  Predictions are given
for the structure function $F_2^{\gamma(P^2)}(x,Q^2)$ and $f^{\gamma(P^2)}
(x,Q^2)$, and are compared with all relevant data for real photons
as well as with recent data for virtual photons as extracted from DIS
$ep$ dijet events.  Simple analytic parametrizations of our predicted
parton distributions are presented for the real photon in LO and NLO,
and for the virtual photon in LO which, within sufficient accuracy, may
be also used in NLO--QCD. 
\end{abstract}
\end{titlepage}
\section{Introduction}
Modern theoretical QCD studies \cite{ref1,ref2,ref3} of the parton 
distributions of {\underline{real}}, i.e.\ \mbox{on--shell}, photons 
$f^{\gamma}(x,Q^2),\,\, f=q,\, \bar{q},\, g$, agree surprisingly well 
with measurements of the (anti) quark and gluon contents of the resolved 
real photons as obtained from $e^+e^-$ and ep reactions at collider 
energies (for recent reviews, see \cite{ref4,ref5,ref6}).  For clarity
let us denote the resolved real target photon with virtuality 
$P^2\equiv -p^2\simeq0$ by $\gamma\equiv\gamma(P^2\simeq 0)$ which is
\mbox{probed} by the virtual probe photon $\gamma^*(Q^2),\, Q^2\equiv -q^2$,
via the subprocess $\gamma^*(Q^2)\gamma\to X$ as in $e^+e^-\to e^+e^-X$.
Here, $p$ denotes the four--momentum of the photon emitted from, say,
an electron in an $e^+e^-$ or ep collider.  In the latter case it is
common to use $Q^2$ instead of $P^2$ for denoting the photon's virtuality,
but we prefer $P^2$ for the subprocess $\gamma(P^2)p\to X$ according
to the original notation used in $e^+e^-$ annihilations.  (Thus the
\mbox{factorization} scale in $f^{\gamma(P^2)}(x,Q^2)$ refers now to some
properly chosen scale of the \mbox{produced} hadronic system $X$, e.g.\ 
$Q\sim p_T^{\rm{jet}}$ in high--$p_T$ jet events, etc.).

In general one expects [7--12] also a {\underline{virtual}}
photon $\gamma(P^2\neq 0)$ to possess a parton content $f^{\gamma(P^2)}
(x,Q^2)$.  It is a major problem to formulate a consistent set of
boundary conditions which allow for a calculation of $f^{\gamma(P^2)}
(x,Q^2)$ also in the next--to--leading order (NLO) of QCD
{\underline{as well as}} for a smooth transition to $P^2=0$,\, i.e.\
to the parton distributions of a real photon (see refs.\ {\cite{ref10,
ref11,ref13} for a detailed discussion).  Indeed, experimental studies
of the transition of the deep inelastic (di--)jet cross section from
the real photon to the virtual photon region at HERA point to the
existence of a nonvanishing, though suppressed, parton content for
virtual photons \cite{ref14}.  These measurements have triggered
various analyses of the dependence of the $ep$ jet production cross
section on the virtuality of the exchanged photon \cite{ref15,ref16}
and experimental tests of such predictions will elucidate the so far 
unanswered question as to when a deep inelastic scattering (DIS) $ep$ 
process is eventually dominated by the usual `direct' $\gamma^*\equiv
\gamma(P^2)$ induced cross sections, {\underline{not}} contaminated
by the so far poorly known {\underline{resolved}} virtual photon
contributions.  More recently, NLO calculations of the (di--)jet
rate in $ep$ (and $e\gamma$) scattering, which properly include the
contributions of resolved virtual photons, have become available
\cite{ref17} and the resolved virtual photon contributions have
already been included in a Monte Carlo event generator \cite{ref18}
as well.

It is the main purpose of the present paper to formulate a consistent
set of boundary conditions, utilizing our valence--like input parton
distributions at the universal target--mass independent 
\cite{ref1,ref19,ref20} low resolution scale $Q_0^2=\mu^2\simeq 0.3$
GeV$^2$, which allow for a calculation of $f^{\gamma(P^2)}(x,Q^2)$ 
also in NLO--QCD as well as for a smooth transition to the parton
distributions of a real photon, $P^2=0$.  We shall furthermore employ
the recently updated parton distributions of the pion \cite{ref21},
$f^{\pi}(x,Q^2)$, which are required for describing, via vector meson
dominance (VMD), the hadronic components of the photon.  It should be
noted that the pionic gluon and sea densities, $g^{\pi}(x,Q^2)$ and
$\bar{q}\,^{\pi}(x,Q^2)$, can be uniquely derived \cite{ref21,ref22} 
from the
experimentally rather well known pionic valence density $v^{\pi}(x,Q^2)$
and the (also recently updated \cite{ref23} dynamical) parton distributions
$f(x,Q^2)$ of the proton.  Thus we arrive at essentially 
parameter--{\underline{free}} predictions for $f^{\gamma(P^2)}(x,Q^2)$
which are furthermore in good agreement with all present measurements
of the structure function of real photons, $F_2^{\gamma}(x,Q^2)$.

In Sec.\ 2 we discuss the basic theoretical framework necessary for
the presentation of our model for the parton distributions and structure
functions of real photons, and \mbox{compare} the resulting predictions with
recent experiments.  Sec.\ 3 contains the formulation of our model
for the parton distributions of virtual photons, together with some
quantitative predictions for structure functions as well as a comparison
with very recent data extracted from DIS dijet events.  Our conclusions  
are drawn in Sec.\ 4.  In the \mbox{Appendix} we present simple analytic
parametrizations of our LO and NLO predictions for the parton 
distributions of real and virtual photons.

\section{The Parton Content of Real Photons}
High energy photons are mainly produced by the bremsstrahlung process
$e(k)\to e(k')$ \mbox{$+\,\gamma(p)$}.  Here 
$p^2=(k'-k)^2=2m_e^2-2k\cdot k'$ 
determines the virtuality of the produced \mbox{photon} which is declared
as real (virtual) whenever $P^2\equiv -p^2$ is smaller (larger) than
some $P_0^2$ arbitrarily fixed experimentally (typically, $P_0^2\simeq
10^{-2}$ GeV$^2$).  Accordingly, the theoretical analysis is usually
subdivided into two distinct parts corresponding to 
\mbox{$P^2$ \raisebox{-0.05cm} {$\stackrel{\textstyle<}{>}$} $P_0^2$}.  
In terms of the photonic parton distributions, 
the structure function $F_2^{\gamma}$ of a real photon 
\mbox{$\gamma\equiv\gamma(P^2$ \raisebox{-0.1cm} 
{$\stackrel{<}{\sim}$} $P_0^2)$} is given in 
NLO($\overline{\rm{MS}}$) QCD by
\begin{eqnarray}
\frac{1}{x}F_2^{\gamma} (x,Q^2) & = & \sum_q e_q^2 \bigg\{  q^{\gamma}
   (x,Q^2)+\bar{q}\,^{\gamma}(x,Q^2)  \nonumber\\
& &  + \, \frac{\alpha_s(Q^2)}{2\pi}\, \left[ C_q\otimes
      (q+\bar{q})^{\gamma} 
  + 2\,C_g\otimes g^{\gamma}\right] +\, \frac{\alpha}{\pi}\, e_q^2 C_{\gamma}
     (x) \bigg\}
\end{eqnarray}
where $\otimes$ denotes the usual convolution integral.  Here 
$\bar{q}\,^{\gamma}(x,Q^2)=q^{\gamma}(x,Q^2)$ and $g^{\gamma}(x,Q^2)$
provide the so--called `resolved' contributions of 
$\gamma$ to $F_2^{\gamma}$, while $C_{\gamma}$ provides the `direct'
contribution as calculated according to the pointlike `box' diagram  
$\gamma^*(Q^2)\gamma(P^2\simeq 0)\to q\bar{q}$.  The Wilson coefficients
\cite{ref24} $C_{q,g}$ are given by
\begin{eqnarray}
C_q(x) & = & \frac{4}{3}\, \left[ \frac{1+x^2}{1-x}\, \left( ln\,
   \frac{1-x}{x}-\frac{3}{4}\right) + \frac{1}{4}\, 
     (9+5x)\right]_+\nonumber\\
C_g(x) & = & \frac{1}{2} \left[ \left(x^2+(1-x)^2\right) ln\, \frac{1-x}{x} 
         + 8x(1-x)-1 \right],
\end{eqnarray}
where the convolution with the $[$ $]_+$ distribution can be easily
calculated using, for example, eq.\ (A.21) of the first article in ref.\ 
\cite{ref19}, while the direct term in (1) is \cite{ref25}
\begin{equation}
C_{\gamma}(x)=\frac{3}{(1/2)}\, C_g(x)=3\, \left[ \left( x^2+(1-x)^2\right)
   \,ln \, \frac{1-x}{x}+8x(1-x)-1 \right].
\end{equation}
The appropriate coefficient function $C_{q,L}$ and $C_{g,L}$ for the 
longitudinal structure function $F_L^{\gamma}=F_2^{\gamma}-2xF_1^{\gamma}$
may be found in \cite{ref10}. 

In order to avoid the usual instabilities encountered in 
NLO($\overline{\rm{MS}}$)
in the large--$x$ region due to the $ln(1-x)$ term in (3), we follow
ref.\ \cite{ref1} and absorb such terms into the photonic 
$\overline{\rm{MS}}$
quark distributions in (1):  this results in the so--called DIS$_{\gamma}$
factorization scheme which originally has been introduced for real
photons by absorbing the entire $C_{\gamma}$ term appearing in (1)
into the NLO($\overline{\rm{MS}})$ quark densities 
$\stackrel{(-)}{q}\!^{\gamma}(x,Q^2)$, i.e.\
\begin{eqnarray}
(q+\bar{q})_{\rm{DIS}_{\gamma}}^{\gamma} & = & 
   (q+\bar{q})_{\overline{\rm{MS}}}^{\gamma} + \frac{\alpha}{\pi}\,
       e_q^2\, C_{\gamma}(x)\nonumber\\
 g_{\rm{DIS}_{\gamma}}^{\gamma} & = & g_{\overline{\rm{MS}}}^{\gamma}
\end{eqnarray}
with $C_{\gamma}(x)$ given by eq.\ (3).  How much of the `finite' terms
in (3) is absorbed into the $\overline{\rm{MS}}$ distributions in (4),
is of course arbitrary and a matter of convention \cite{ref3}.  Since
such different conventions \cite{ref3,ref5} turn out to be of minor
importance for our quantitative results to be discussed below, we prefer
to stick to the original DIS$_{\gamma}$ scheme \cite{ref1} as defined
in eq.\ (4).  Furthermore, the redefinition of the parton densities in
(4) imply that the NLO($\overline{\rm{MS}}$) splitting
functions $k_{q,g}^{(1)}(x)$ of the photon into quarks and gluons,
appearing in the inhomogeneous NLO RG evolution equations \cite{ref1,ref10}
for $f^{\gamma}(x,Q^2)$, have to be transformed according to 
\cite{ref1,ref26}
\begin{eqnarray}
k_q^{(1)}|_{\rm{DIS}_{\gamma}} & = & k_q^{(1)}|_{\overline{\rm{MS}}} -
		e_q^2\, P_{qq}^{(0)} \otimes C_{\gamma}\nonumber\\
k_g^{(1)}|_{\rm{DIS}_{\gamma}} & = & k_g^{(1)}|_{\overline{\rm{MS}}} -
		2\, \sum_q  e_q^2\, P_{gq}^{(0)}\otimes C_{\gamma}\, .
\end{eqnarray}

The LO expression for $F_2^{\gamma}$ is obviously entailed in eq.\ (1)
by simply dropping all NLO terms proportional to $C_{q,g}$ as well as
$C_{\gamma}$.  

In NLO the expression for $F_2^{\gamma}$ in the above DIS$_{\gamma}$
factorization scheme is given by retaining the $C_{q,g}$ terms while
dropping the destabilizing $C_{\gamma}$ terms in eq.\ (1), which has
already been absorbed into the quark densities according to eq.\ (4).
Furthermore the nonperturbative hadronic VMD input $f^{\gamma}(x,Q_0^2)$
in NLO refers to the partons in the DIS$_{\gamma}$ scheme which guarantees
the perturbative stability of the resulting $F_2^{\gamma}(x,Q^2)$ provided
this input is given by the NLO $f^{\pi}(x,Q_0^2)$ of \cite{ref21}, while
the corresponding input in LO is given via VMD by the LO  
$f^{\pi}(x,Q_0^2)$ of \cite{ref21}.
We shall assume that the input resolution scale 
$Q_0^2=\mu^2\simeq 0.3$ GeV$^2$ for the valence--like parton structure
is universal, i.e. independent of the mass of the considered targets
$p,\, \pi, \, \gamma$, etc. \cite{ref1,ref19,ref20,ref22}.  The hadronic
VMD ansatz for $f^{\gamma}(x,\mu^2)$ is based on a coherent superposition
of vector mesons \cite{ref11}
\begin{equation}
|\gamma\rangle_{\mu^2,\,\rm{had}}\simeq \frac{e}{f_\rho}\, 
  |\rho\rangle_{\mu^2} + \frac{e}f_\omega\, |\omega\rangle_{\mu^2}
\end{equation} 
where the $\phi$--meson contribution is considered to be strongly 
suppressed at $\mu^2\ll m_{\phi}^2$.  Assuming, within the operator
product expansion (OPE),
\begin{eqnarray}
\langle\rho|{\rm{O}}_q|\rho\rangle_{\mu^2} & = & \langle\omega|{\rm{O}}_q|
  \omega\rangle_{\mu^2} = 2\, I_{3q}\, e^{-i\theta} \langle\rho|{\rm{O}}_q|
    \omega\rangle_{\mu^2} = \langle\pi^0|{\rm{O}}_q|\pi^0\rangle_{\mu^2}\\
\langle\rho|{\rm{O}}_g|\rho\rangle_{\mu^2} & = & \langle\omega|{\rm{O}}_g|
   \omega\rangle_{\mu^2} = \langle\pi^0|{\rm{O}}_g|\pi^0\rangle_{\mu^2}
\end{eqnarray}
and $\langle\rho|{\rm{O}}_g|\omega\rangle=0$ due to isospin conservation, 
one obtains,
\begin{eqnarray}
(u+\bar{u})^{\gamma}(x,\mu^2) & = & \alpha(g_{\rho}^2 + g_{\omega}^2 +
  2\,g_{\rho}g_{\omega}\cos\theta)(u+\bar{u})^{\pi^0}(x,\mu^2)\nonumber\\
(d+\bar{d})^{\gamma}(x,\mu^2) & = & \alpha(g_{\rho}^2 + g_{\omega}^2 -
  2\,g_{\rho}g_{\omega}\cos\theta)(d+\bar{d})^{\pi^0}(x,\mu^2)\nonumber\\
(s+\bar{s})^{\gamma}(x,\mu^2) & = & \alpha(g_{\rho}^2 + g_{\omega}^2)\,
    (s+\bar{s})^{\pi^0}(x,\mu^2) = 0\nonumber\\
g^{\gamma}(x,\mu^2) & = & \alpha(g_{\rho}^2 + g_{\omega}^2)\, g^{\pi^0}
   (x,\mu^2) .
\end{eqnarray}
Here ${\rm{O}}_{q,g}$ refer to the leading twist--2 quark and gluon 
operators in the OPE formalism, $g^{\gamma}\equiv\langle\gamma|{\rm{O}}_g|
\gamma\rangle_{\rm{had}}\,$ etc., and $g_V^2\equiv 4\pi/f_V^2$ with
\begin{equation}
g_{\rho}^2 = 0.50\, , \quad\quad g_{\omega}^2 = 0.043\, ,
\end{equation}
i.e.\ $f_{\rho}^2/4\pi = 2.0$ and $f_{\omega}^2/4\pi =23.26$, as obtained 
from a zero--width calculation of the relevant leptonic widths
$\Gamma(V\to\ell^+\ell^-)=\alpha^2m_V g_V^2/3$ presented in \cite{ref27}.
The omission of a finite--width correction for $g_{\rho}^2$ is due to
the central role \cite{ref27} of the precise results in \cite{ref28} which
do not require such a correction in contrast to the situation for the
less precise resonance analysis at $e^+e^-$ colliders \cite{ref29}.

For the a priori unknown coherence factor (fit parameter) $\cos\theta$
in eq.\ (9) we take $\cos\theta=1$, i.e.\ we favor a superposition of
$u$ and $d$ quarks  which maximally enhances the contributions of the
up--quarks to $F_2^{\gamma}$ in eq.\ (1).  This favored value for 
$\cos\theta$ is also supported by fitting $\cos\theta$ in (9) to all
presently available data on $F_2^{\gamma}(x,Q^2)$, to be discussed 
below, which always resulted in $\cos\theta\simeq 1$ in LO as
well as NLO.  This is also in agreement with the LO results obtained
in ref.\ \cite{ref11}.  The LO and NLO input distributions $f^{\pi}
(x,\mu^2)$ of the pion in (9) are taken from a recent analysis
\cite{ref21} which correspond to \cite{ref21,ref23} $\mu_{\rm{LO}}^2 =
0.26$ GeV$^2$ and $\mu_{\rm{NLO}}^2=0.40$ GeV$^2$ in LO and NLO, 
respectively.  Since by now all free input quantities have been fixed
in eq.\ (9), we arrive at rather unique parameter--{\underline{free}}
predictions for $f^{\gamma}(x,Q^2)$ and $F_2^{\gamma}(x,Q^2)$.

The calculation of $f^{\gamma}(x,Q^2)$ at $Q^2>\mu^2$ follows from the
well known inhomogeneous RG evolution equations in LO and NLO (see,
for example, \cite{ref1,ref26}) which we solve, as usual, analytically
for the $n$--th Mellin moment of $f^{\gamma}(x,Q^2)$, followed by a 
straightforward Mellin--inversion to Bjorken--$x$ space.  The explicit
formal solutions can be found in eqs.\ (2.12) and (2.13) of the first
article of ref.\ \cite{ref1}.  The general structure of these solutions
is 
\begin{equation}
f^{\gamma}(x,Q^2) = f_{pl}^{\gamma}(x,Q^2) + f_{\rm{had}}^{\gamma} 
  (x,Q^2)\, .
\end{equation}
Here $f_{pl}^{\gamma}$ denotes the perturbative `pointlike' solution
which vanishes at $Q^2=\mu^2$ and is driven by the pointlike photon
splitting functions $k_{q,g}^{(0,1)}(x)$ appearing in the inhomogeneous
evolution equations, while $f_{\rm{had}}^{\gamma}$ depends on the hadronic
input $f^{\gamma}(x,\mu^2)$ in eq.\ (9) and evolves according to the
standard homogeneous evolution equations.  We treat these solutions
in precisely the same way as discussed in detail in \cite{ref10}, except 
that we
implement an improved treatment of the running $\alpha_s(Q^2)$ by exactly
solving in NLO($\overline{\rm{MS}}$)
\begin{equation}
\frac{d\,\alpha_s(Q^2)}{d\,ln\,Q^2} = -\frac{\beta_0}{4\pi}\, \alpha_s^2(Q^2)
   - \frac{\beta_1}{16\pi^2}\, \alpha_s^3(Q^2)
\end{equation}
numerically \cite{ref23} using $\alpha_s(M_Z^2)= 0.114$, rather than using
the usual approximate NLO solution (e.g.\ eq.\ (14) in \cite{ref10}) 
which becomes sufficiently accurate only for $Q^2 $\raisebox{-0.1cm}
{$\stackrel{>}{\sim}$} $m_c^2 \simeq$ \mbox{2 GeV$^2$} \cite{ref23}.  Here,
$\beta_0=11-2f/3$ and $\beta_1=102-38f/3$.

The prescription for the VMD ansatz in eq.\ (9) at the input scale
$\mu^2$, together with $\cos\theta=1$ as discussed above, yields a
simple expression for the general $Q^2$--dependence of $f^{\gamma}(x,Q^2)$:
\begin{equation}
f^{\gamma}(x,Q^2) = f_{pl}^{\gamma}(x,Q^2)+\alpha\, \left[G_f^2\, f^{\pi}(x,Q^2)+
\delta_f\, \frac{1}{2}(G_u^2-G_d^2)\, s^{\pi}(x,Q^2)\right]
\end{equation}
with $\delta_u=-1$, $\delta_d=+1$ and $\delta_s=\delta_g=0$, and
where the index $\pi$ obviously refers to $\pi^0$ and 
\newpage

\begin{eqnarray}
G_u^2 & = & (g_{\rho}+g_{\omega})^2 \simeq 0.836\nonumber\\
G_d^2 & = & (g_{\rho}-g_{\omega})^2 \simeq 0.250\nonumber\\
G_s^2 & = & G_g^2 = g_{\rho}^2 + g_{\omega}^2 = 0.543\, .
\end{eqnarray}
Simple analytic LO and NLO(DIS$_{\gamma}$) parametrizations for the
pointlike piece $f_{pl}^{\gamma}(x,Q^2)$ are given in the Appendix,
whereas the ones for $f^{\pi}(x,Q^2)$ can be found in \cite{ref21}.  

The photonic quark distributions discussed thus far and which appear
in eq.\ (1) are adequate for the $f=3$ light $u,\, d,\, s$ flavors.
Since heavy quarks $h=c,\,b,\,t$ will not be considered as `light'
partons in the photon (as in the case of the proton \cite{ref23} and the
pion \cite{ref21,ref22}), their contributions to $F_2^{\gamma}$ have
to be calculated in fixed order perturbation theory according to the
`direct' box--diagram $\gamma^*(Q^2)\,\gamma\to h\bar{h}$ expression, i.e.\
the usual Bethe--Heitler cross section \cite{ref30}
\begin{eqnarray}
\frac{1}{x}\,F_{2,h}^{\gamma}(x,Q^2) & = & 3\, e_h^4\, \frac{\alpha}{\pi}\,
   \theta(\beta^2)\,\left\{ \beta\,\left[ 8x(1-x)-1-x(1-x) \,
     \frac{4m_h^2}{Q^2}\right] \right. \nonumber\\
& & \left. + \left[ x^2+(1-x)^2 + x(1-3x)\, \frac{4m_h^2}{Q^2}-x^2\, 
        \frac{8m_h^4}{Q^4}\right] \, ln\, \frac{1+\beta}{1-\beta}\, \right\}
\end{eqnarray}
where $\beta^2\equiv1-4m_h^2/W^2=1-4m_h^2x/(1-x)Q^2$.  This expression
has to be added to eq.\ (1) and a similar expression holds for the 
longitudinal structure function \cite{ref10} $F_L\equiv F_2-2xF_1$.  
The `resolved' 
heavy quark contribution \cite{ref10} to $F_2^{\gamma}$ in (1) has to be
calculated via \mbox{$\gamma^*(Q^2)g^{\gamma}\to h\bar{h}$},
\begin{equation}
F_{2,h}^{g^{\gamma}}(x,Q^2) = \int_{z_{min}}^1\frac{dz}{z}\, zg^{\gamma}
  (z,\mu_F^2)\, f_2^{\gamma^*(Q^2)g^{\gamma}\to h\bar{h}}\left( 
           \frac{x}{z},\,Q^2 \right)
\end{equation}
where $\frac{1}{x}\, f_2^{\gamma^*(Q^2)g^{\gamma}\to h\bar{h}}(x,Q^2)$
is given by eq.\ (15) with $e_h^4\alpha\to e_h^2\alpha_s(\mu_F^2)/6$,
$z_{\min}=$ \mbox{$x(1+ 4m_h^2/Q^2)$} and $\mu_F^2\simeq 4m_h^2$ 
\cite{ref31}.
This `resolved' LO contribution should be included in eq.\ (1) as well.
To ease the calculations we shall keep these LO expressions (15) and
(16) also in NLO, since the full NLO expressions for heavy quark 
production \cite{ref32} turn out to be a small correction to the 
already not too sizeable (at most about 20\%) contribution in LO.
Notice that such small corrections are not larger than ambiguities
due to different choices for $m_h$ and for the factorization scale
$\mu_F$.  For our purposes it is sufficient to include only the charm
contributions which will be calculated using $m_c=1.4$ GeV.

Having outlined the theoretical basis for our photonic parton 
distributions, we now turn to the quantitative results.
First we apply our parameter--free predictions for $f^{\gamma}(x,Q^2)$
to the structure function of real photons which, according to
eq.\ (1) and the above results, is finally given by
\begin{eqnarray}
\frac{1}{x}\, F_2^{\gamma}(x,Q^2) & = & 2\sum_{q=u,d,s} e_q^2 \left\{ 
  q^{\gamma}(x,Q^2)+\,\frac{\alpha_s(Q^2)}{2\pi}\, \left[ C_q\otimes 
    q^{\gamma} + C_g\otimes g^{\gamma} \right] \right.\nonumber\\
& &  \left. +\, \frac{1}{x}\, F_{2,c}^{\gamma}(x,Q^2) + \, \frac{1}{x}\, 
     F_{2,c}^{g{^\gamma}}(x,Q^2) \right\}
\end{eqnarray}
where $f^{\gamma}(x,Q^2)$ refers to the DIS$_{\gamma}$ factorization
scheme defined in (4) and the charm contributions $F_{2,c}^{\gamma(P^2)}$
and $F_{2,c}^{g^{\gamma}}$ are given by eqs.\ (15) and (16), respectively.
In fig.\ 1 we compare our LO and NLO predictions with all available 
relevant data \cite{ref33} for $F_2^{\gamma}$ of the real photon. Our
present new NLO results are rather similar to the ones of AFG \cite{ref3},
but differ from our previous (GRV$_{\gamma}$) predictions \cite{ref1}
which are steeper in the small--$x$ region, as shown in fig.\ 1, because
the dominant hadronic (pionic) sea density $\bar{q}\,^{\pi}(x,Q^2)$ is
steeper since it has been generated purely dynamically from a vanishing
input at $Q^2=\mu^2$ \cite{ref1,ref20}.  Similarly the SaS 1D \cite{ref11}
expectations, for example, fall systematically below the data in the small
to medium $Q^2$ region around $Q^2\simeq 5$ GeV$^2$, partly due to a 
somewhat different treatment of the hadronic coherent VMD input as compared
to our results in eqs.\ (9) and (13) and (14).  The relevant LO and NLO 
photonic parton densities are compared in fig.\ 2 at $Q^2=10$ GeV$^2$.  For
illustration we also show the purely `hadronic' component (homogeneous
solution) in (11) of $f^{\gamma}$ which demonstrates the dominance of 
the `pointlike' component (inhomogeneous solution) in (11) for 
$u^{\gamma}$ and $d^{\gamma}$ in the large--$x$ region, $x>0.1\,$.  
In fig.\ 3  we show our predictions for $xu^{\gamma}(x,Q^2)$ and 
$xg^{\gamma}(x,Q^2)$.  The parton distributions of the photon behave,
in contrast to the ones of a hadron, very differently in the limits
of large and small $x$.  In the former case, the purely perturbative
pointlike part in (11) dominates for $x$ \raisebox{-0.1cm}
{$\stackrel{>}{\sim}$} 0.1, especially for the quark distributions.
On the other hand, this uniquely calculable contribution amounts at
most to about 20\% at very small $x$ where the hadronic VMD component
in (11) dominates, giving rise to a very similar increase for $x\to 0$
as observed in the proton case.  In fig.\ 3 we also show our valence--like
inputs at $Q^2=\mu^2_{\rm{LO,\,NLO}}$ which become (vanishingly) small
at $x<10^{-2}$.  This illustrates the purely dynamical origin of the
small--$x$ increase at $Q^2>\mu^2$.  Also noteworthy is the perturbative
LO/NLO stability of $u^{\gamma}(x,Q^2)$ which is almost as good as the
one required for a physical quantity like $F_2^{\gamma}(x,Q^2)$ in fig.\ 1.
The situation is, as usual \cite{ref1,ref19,ref23}, different for
$g^{\gamma}(x,Q^2)$.  Nevertheless, despite the sizable difference 
between the LO and NLO gluon distributions in fig.\ 3 in the small--$x$
region, the directly measurable $F_2^{\gamma}$ and the gluon--dominated          
heavy quark contribution in eq.\ (16) shows a remarkable perturbative
stability \cite{ref23}.  Finally, we compare in fig.\ 4 our predictions
for $xg^{\gamma}(x,Q^2)$ at \mbox{$Q^2\equiv(p_T^{\rm{jet}})^2=75$ GeV$^2$} 
with recent HERA (H1) measurements \cite{ref34}.  Our somewhat flatter 
results for $xg^{\gamma}$ in the small--$x$ region, as compared to the
older GRV$_{\gamma}$ expectations \cite{ref1}, is caused by the recently
favored flatter gluon distribution in the proton \cite{ref23} which
determines $g^{\gamma}$ via $g^{\pi}$ \cite{ref21}, \mbox{cf.\ eq.\ (9)},
at small $x$.

Finally it is interesting to consider the total momenta carried
by the photonic partons,
\begin{equation}
M_2^{\gamma}(Q^2)\equiv \sum_{f=q,\bar{q},g} \int_0^1x\, f^{\gamma}
  (x,Q^2)\, dx\, .
\end{equation}
Inspired by the ideas and suggestions put forward in refs.\ 
\cite{ref35,ref11}, it has been conjectured recently \cite{ref36} that
this leading twist--2 quantity $M_2^{\gamma}$ should satisfy, in 
LO-QCD,
\begin{equation}
M_2^{\gamma}(Q^2)\simeq \Pi_h(Q^2)
\end{equation}
where the well known dispersion relation relates the hadronic part of
the photon's vacuum polarization
\begin{equation}
\Pi_h(Q^2)=\frac{Q^2}{4\pi^2\alpha}\, \int_{4m_{\pi}^2}^{\infty}\,
   \frac{\sigma_h(s)}{s+Q^2}\, ds
\end{equation}
to $\sigma_h\equiv\sigma(e^+e^-\to$ hadrons).  It should be noted that
$\Pi_h(Q^2)$, being an experimental quantity, includes, besides the
usual twist--2 term, all possible nonperturbative higher--twist 
contributions.  The `consistency' relation (19) is, however, expected
to hold already at $Q^2$ \raisebox{-0.1cm}{$\stackrel{>}{\sim}$} 2 to
4 GeV$^2$ to within, say, 20 to 30\% where the twist--2 component
in $\Pi_h(Q^2)$ may become dominant, as possibly indicated by DIS $ep$
processes.  Indeed, our LO results imply 
$M_2^{\gamma}(2\,\,{\rm{GeV}}^2)/\alpha\simeq 0.976$ 
and $M_2^{\gamma}(4\,\,{\rm{GeV}}^2)/\alpha\simeq 1.123$
which compares favorably with \cite{ref37} $\Pi_h(2\,\,{\rm{GeV}}^2)/
\alpha=0.694\pm0.028$ and $\Pi_h(4\,\,{\rm{GeV}}^2)/\alpha=0.894\pm0.036$,
respectively.

%
\section{The Parton Content of Virtual Photons}
Next we turn to the somewhat more speculative concept and models of
`resolved' virtual photons ($P^2\neq 0$). The real photons considered
in Section 2 are those whose virtuality $P^2$ is very small, i.e.\ of
the order $P_{min}^2={\cal{O}}(m_e^2)$ or, experimentally, at least 
$P^2<P_0^2\simeq 10^{-2}$ GeV$^2$.  
The flux of {\underline{virtual}} photons
produced by the bremsstrahlung process $e(k)\to e(k')+\gamma(p)$,
$P^2\equiv -p^2 = -(k'-k)^2 > P_0^2$ is given by \cite{ref38}
\begin{eqnarray}
f_{\gamma(P^2)/e}^T(y) & = & \frac{\alpha}{2\pi} \, 
  \left[ \frac{1+(1-y)^2}{y}\, \frac{1}{P^2}\, -\, \frac{2m_e^2\, y}
    {P^4} \right]\\
f_{\gamma(P^2)/e}^L(y) & = & \frac{\alpha}{2\pi} \, \frac{2(1-y)}{y}
       \, \frac{1}{P^2}
\end{eqnarray}
with $y=E_{\gamma}/E_e$ and $T(L)$ denoting transverse (longitudinal)
photons.  Whenever these virtual photons, with their virtuality being
entirely taken care of by the flux factors in (21) and (22), are probed 
at a scale $Q^2\gg P^2$ they may be considered as {\underline{real}}
photons which means that \cite{ref7,ref8,ref12,ref15,ref16,ref17}
\vspace{-0.5cm}
\begin{itemize}
\item[(i)] effects due to $f_{\gamma(P^2)/e}^L$ should be neglected
	since the corresponding longitudinal cross sections are
	suppressed by powers of $P^2/Q^2$;
\item[(ii)] cross sections of partonic subprocesses involving 
	$\gamma(P^2)$ should be calculated as if $P^2=0$ due (partly)
	to the $P^2/Q^2$ power suppressions of any additional terms.
\end{itemize}
This latter rule implies in particular that the NLO `direct' contribution
$C_{\gamma(P^2)}(x)$ to $F_2^{\gamma(P^2)}(x,Q^2)$ has to be the 
{\underline{same}} $C_{\gamma}(x)$ as for {\underline{real}} photons
in eqs.\ (1) and (3), i.e.\ has to be inferred from the real photon
subprocess $\gamma^*(Q^2)\gamma\to q\bar{q}$, and {\underline{not}} 
from the doubly--virtual box $\gamma^*(Q^2)\gamma(P^2)\to q\bar{q}$ as
originally proposed \cite{ref7,ref8} and used \cite{ref10}.  Thus we
can implement the same DIS$_{\gamma}$ factorization scheme as for real
photons in eq.\ (4), and the structure function of virtual photons
becomes formally very similar to eq.\ (17):
\begin{eqnarray}
\frac{1}{x}\, F_2^{\gamma(P^2)}(x,Q^2) & = & 2\sum_{q=u,d,s} e_q^2 \left\{ 
  q^{\gamma(P^2)}(x,Q^2)+\,\frac{\alpha_s(Q^2)}{2\pi}\, \left[ C_q\otimes 
    q^{\gamma(P^2)} + C_g\otimes g^{\gamma(P^2)} \right] \right.\nonumber\\
& &  \left. +\, \frac{1}{x}\, F_{2,c}^{\gamma}(x,Q^2) + \, \frac{1}{x}\, 
     F_{2,c}^{g{^{\gamma(P^2)}}}(x,Q^2) \right\}
\end{eqnarray}
with $C_{q,g}(x)$ given in (2).  The `direct' heavy (charm) quark 
contribution is given by \mbox{eq.\ (15)} as for real photons since 
$F_{2,c}^{\gamma(P^2)}=F_{2,c}^{\gamma}$ as follows from our consistent
strict adherence to point (ii) above.  The  `resolved' charm contribution
$F_{2,c}^{g^{\gamma(P^2)}}$ is as in \mbox{eq.\ (16)} with the gluon 
distribution $g^{\gamma}(z,\mu_F^2)\to g^{\gamma(P^2)}(z,\mu_F^2)$.

The above consistency requirements afford furthermore the following
boundary conditions for $f^{\gamma(P^2)}$, cf.\ eq.\ (13),
\begin{equation}
f^{\gamma(P^2)}(x,Q^2=\tilde{P}^2) = 
   f_{\rm{had}}^{\gamma(P^2)}(x,\tilde{P}^2) =
     \eta(P^2) f_{\rm{had}}^{\gamma}(x,\tilde{P}^2)
\end{equation}
in LO {\underline{as well as}} in NLO.  Here $\tilde{P}^2={\rm{max}}\,
(P^2,\mu^2)$ as dictated by continuity in $P^2$ \cite{ref10} and 
$\eta(P^2)=(1+P^2/m_{\rho}^2)^{-2}$ is a dipole suppression factor with
$m_{\rho}^2=0.59$ GeV$^2$.  The second equality in eq.\ (24) follows 
from the consistency requirement $C_{\gamma(P^2)}=C_{\gamma}$ and
consequently the application of the {\underline{same}} $\overline{\rm{MS}}
\to DIS_{\gamma}$ factorization scheme transformation as for the 
{\underline{real}} photon, cf.\ eq.\ (23).  The scale $\tilde{P}^2$
is dictated not only by the above mentioned continuity requirement, but 
also by the fact that the hadronic component of $f^{\gamma(P^2)}(x,Q^2)$
is probed at the scale $Q^2=\tilde{P}^2$ \cite{ref9,ref10,ref11} where
the pointlike component vanishes by definition.  The boundary condition
in eq.\ (24) guarantees, as should be evident, a far better perturbative
stability as compared to the situation in \cite{ref10} where the NLO
input differed drastically from its LO counterpart (cf.\ eq.\ (8) in
ref.\ \cite{ref10}).  

The evolution to $Q^2>\tilde{P}^2$ is now analogous to the case of 
real photons in the previous section and the general solution for the
resulting parton distributions is similar to the one in eq.\ (11) and
eq.\ (13),
\begin{eqnarray}
f^{\gamma(P^2)}(x,Q^2) & = & f_{pl}^{\gamma(P^2)}(x,Q^2) + 
    f_{\rm{had}}^{\gamma(P^2)}(x,Q^2)\\
 & =  & f_{pl}^{\gamma(P^2)}(x,Q^2)+\eta(P^2)\alpha \left[G_f^2\, f^{\pi}(x,Q^2)+
\delta_f\, \frac{1}{2}(G_u^2-G_d^2)\, s^{\pi}(x,Q^2)\right]\nonumber
\end{eqnarray}
with $\delta_f$ as in eq.\ (13) and
where $f_{\rm{had}}^{\gamma(P^2)}$ refers again to the solution of the
homogeneous RG evolution equations, being driven by the hadronic input
in (24), which is explicitly given by eq.\ (13) of ref. \cite{ref10}.
\mbox{Its parametrization} is fixed by the available parametrization 
\cite{ref21} for $f^{\pi}(x,Q^2)$ in (25).  The inhomogeneous `pointlike' 
solution in (25) is explicitly given by eq.\ (12) of \cite{ref10} 
where $L=\alpha_s(Q^2)/\alpha_s(\tilde{P}^2)$.  A parametrization of 
$f_{pl}^{\gamma(P^2)}(x,Q^2)$ in LO is thus easily obtained from the 
one for the real photon $f_{pl}^{\gamma}(x,Q^2)$ in (13) in terms of 
$ln \, L^{-1}=ln\,\left[\alpha_s(\mu^2)/\alpha_s(Q^2)\right]$, where now 
$\alpha_s(\mu^2)$ has simply to be replaced by $\alpha_s(\tilde{P}^2)$ 
as described in detail in Appendix~1.
Furthermore, since our NLO predictions for $f^{\gamma(P^2)}(x,Q^2)$
turn out to be rather similar to the LO ones, as will be shown below,
the simple analytic LO parametrizations for $f^{\gamma(P^2)}(x,Q^2)$
can be used for NLO calculations as well.  This is certainly sufficiently
accurate and reliable in view of additional model ambiguities inherent
in the parton distributions of virtual photons.

It should be emphasized that the RG resummed results in (25) are 
relevant when-\- ever $P^2\ll Q^2$, typically \cite{ref10,ref15} 
$P^2\simeq\frac{1}{10}\,\,Q^2$, so as to suppress power--like (possibly 
higher twist) terms $(P^2/Q^2)^n$ which would spoil the dominance of 
the resummed logarithmic contributions and, furthermore,  to guarantee
the dominance of the transverse photon contributions (21) to physical
cross sections.  For $P^2$ approaching $Q^2$, the $e^+e^-\to e^+e^- X$
reaction, for example, should be simply described by the full fixed
order box $\gamma^*(Q^2)\gamma(P^2)\to q\bar{q}$ keeping all $(P^2/Q^2)^n$
terms.  Since the full perturbative ${\cal{O}}(\alpha_s)$ corrections
to this virtual box have not been calculated yet, it is not possible,
for the time being, to determine reliably at what values of $P^2$
(and possibly $x$) this ${\cal{O}}(\alpha_s)$ 
corrected virtual box becomes the more appropriate and correct description.
Similar remarks hold for a DIS process $ep\to eX$, i.e. 
$\gamma(P^2)\, p\to X$, where ${\cal{O}}(\alpha_s)$ corrections to pointlike
virtual $\gamma(P^2)$--parton subprocesses have to be analyzed in 
detail in order to decide at what $P^2$ these pointlike expressions
become the more appropriate description and the virtual photonic
parton distributions (i.e., resummations) become irrelevant.  

Our strict adherence to the above point (ii) implies that the `direct'
photon contribution to any process whatsoever should always be calculated
as if this photon is real apart from the fact that its flux should be
evaluated according to eq.\ (21) with $P^2\neq 0$ \cite{ref12,ref15,ref16}.
This differs from the somewhat inconsistent procedure adopted by the 
\mbox{HERA--H1} collaboration \cite{ref14} where exact $e\,q\to e\,q\,g$ 
and $e\,g\to e\,q\,\bar{q}$ matrix elements were used for the direct 
photon contribution to the dijet cross section.  As long as 
\mbox{$P^2$ \raisebox{-0.1cm}{$\stackrel
{<}{\sim}$} $\frac{1}{10}\,Q^2$,} the exact treatment of matrix elements,
however, should not differ too much \cite{ref17} from the more appropriate
treatment described above.  To conclude let us stress that the strict
adherence to point (ii), as illustrated by the foregoing examples,
is not a free option but a {\underline{necessary consistency condition}}
for introducing the concept of the resolved parton content of the 
virtual photon as an {\underline{alternative}} to a non--resummed fixed
order perturbative analysis at $P^2\neq 0$.  This consistency requirement
is related to the fact that all the resolved contributions due to
$f^{\gamma(P^2)}(x,Q^2)$ are calculated (evoluted) as if these partons
are massless $[7-11]$ (i.e.\ employing photon splitting
functions for real photons, etc.) in spite of the fact that their actual
virtuality is given by $P^2\neq 0$.  Thus the direct photon contribution
should obviously be also treated accordingly.

Now we turn to our quantitative predictions and first compare our 
results with the old PLUTO measurements \cite{ref39} for 
$F_{\rm{eff}}^{\gamma(P^2)}(x,Q^2)\equiv F_2^{\gamma(P^2)} +\frac{3}{2}
F_L^{\gamma(P^2)}$ in fig.\ 5. Our very similar LO and NLO results, 
which are dominated by the almost unique `pointlike' contribution in
(25), are in full agreement with the limited poor statistics of the 
PLUTO data.  The GRS \cite{ref10} expectations turn out to be very
similar to our present ones shown in fig.\ 5.  
For illustration the naive (i.e.\ not
resummed) LO quark--parton model `box' expectation is shown as well
by the dotted curve in fig.\ 5 which is given by
\begin{eqnarray}
\frac{1}{x}\, F_{2,\rm{box}}^{\gamma(P^2)}(x,Q^2) & = & 3 \sum_{q=u,d,s}
       e_q^4\, \frac{\alpha}{\pi}\, \left[ x^2+(1-x)^2\right]\, ln\,
        \frac{Q^2}{P^2} + \frac{1}{x}\, F_{2,c}^{\gamma(P^2)}(x,Q^2)\\
\frac{1}{x}\, F_{L,{\rm{box}}}^{\gamma(P^2)}(x,Q^2) & = & 3 \sum_{q=u,d,s}
       e_q^4\, \frac{\alpha}{\pi}\, 4x(1-x) + \frac{1}{x}\,
               F_{L,c}^{\gamma(P^2)}(x,Q^2)
\end{eqnarray}
with the heavy quark (charm) contribution $F_{2,c}^{\gamma(P^2)}$ given
by eq.\ (15) and
\begin{equation}
\frac{1}{x}\, F_{L,c}^{\gamma(P^2)}(x,Q^2) = 3\, e_c^4\,\frac{4\alpha}{\pi} 
   \, \left[\beta x(1-x)-x^2\,\frac{2m_c^2}{Q^2}\, ln\,
            \frac{1+\beta}{1-\beta} \right]\, .
\end{equation}
More detailed predictions for $F_2^{\gamma(P^2)}(x,Q^2)$ are presented
in figs.\ 6a and 6b for various virtualities $P^2$ and scales $Q^2$.
Since the `pointlike' component in (25) is uniquely calculable 
perturbatively, a detailed measurement of the $x$ and $P^2$ dependence
at various fixed values of $Q^2$, as shown in figs.\ 6a and 6b, would
shed light on the theoretically more speculative and far less understood
nonperturbative `hadronic' contribution in eq.\ (25) and eventually
establish the absolute perturbative predictions.  Our LO and NLO
predictions in figs.\ 6a and 6b show a remarkable perturbative stability
throughout the whole $x$--region shown, except perhaps for $P^2\gg 1$
GeV$^2$ where the perturbatively very stable \cite{ref21,ref22} `hadronic'
component in (25) becomes strongly suppressed with respect to the 
`pointlike' solution which is less stable in the small $x$ region,
$x<10^{-2}$, as is evident from fig.\ 6b.

The individual LO and NLO parton distributions of the virtual photon
at $Q^2=10$ GeV$^2$ are shown in fig.\ 7a where they are compared with
the ones of GRS \cite{ref10}.  The LO SaS expectations \cite{ref11}
are compared with our LO predictions in fig.\ 7b.  In figs.\ 8 and 9
we show our predictions for $xu^{\gamma(P^2)}(x,Q^2)$ and
$xg^{\gamma(P^2)}(x,Q^2)$ with particular emphasis on the very small
$x$  region.  For comparison we also show the results for a real
$(P^2=0)$ photon.  Plotting the `hadronic' component in (25) separately
in fig.\ 8 demonstrates that the perturbative `pointlike' component in
(25) dominates for $x>10^{-2}$.  Furthermore the expected perturbative
stability of our present LO and NLO predictions is fulfilled.  This
is in contrast to the GRS results which are unstable \cite{ref10}
throughout the whole \mbox{$x$--region} for $P^2$ \raisebox{-0.1cm}
{$\stackrel{>}{\sim}$} 1 GeV$^2$, as illustrated in fig.\ 9 at 
$Q^2=100$ GeV$^2$,
due to the very different perturbative (box) input in LO and NLO 
\cite{ref10}.  In general, however, as soon as the perturbatively
very stable `hadronic' component in (25) becomes suppressed for 
$P^2\gg 1$ GeV$^2$, the remaining perturbatively less stable `pointlike'
component destabilizes the total results for $q^{\gamma(P^2)}(x,Q^2)$ 
in the 
very small $x$ region, $x$ \raisebox{-0.1cm}{$\stackrel{<}{\sim}$}
$10^{-3}$, as can be seen in fig.\ 9 for $u^{\gamma(P^2)}$ at 
$Q^2=100$ GeV$^2$ (cf.\ fig.\ 6b).

Finally in fig.\ 10 we confront our LO predictions for $f^{\gamma(P^2)}
(x,Q^2)$ with the effective parton density
\begin{equation}
\tilde{f}\,^{\gamma(P^2)}(x,Q^2) = \sum_{q=u,d,s}\left(q^{\gamma(P^2)}
   + \bar{q}\,^{\gamma(P^2)}\right) +\, \frac{9}{4}\,g^{\gamma(P^2)}
\end{equation}
extracted in LO from DIS dijet data by the HERA--H1 collaboration
\cite{ref14} very recently.  The predicted dependence on the photon's
virtuality $P^2$ at the scale $Q^2\equiv\left( p_T^{\rm{jet}}\right)^2=85$
GeV$^2$ agrees reasonably well with the measurements in the relevant
kinematic region $P^2\ll Q^2$.  This is also the case at other scales
$Q^2\equiv\left( p_T^{\rm{jet}}\right)^2$ and fixed values of $x$ 
\cite{ref14} not shown in fig.\ 10.  As discussed above, it should be
kept in mind, however, that for larger values of $P^2$ approaching $Q^2$,
which refer to the dashed curves in fig.\ 10, the whole concept of RG
resummed parton distributions of virtual photons is {\underline{not}}
appropriate any-\- more.  Since the resolved contributions of a virtual
photon with virtuality as large as $P^2=10-15$ GeV$^2$ are by a factor
of about 10 smaller than the ones of a real ($P^2=0$) photon, it is
reasonable to conclude from fig.\ 10 that for $P^2$ \raisebox{-0.1cm}
{$\stackrel{>}{\sim}$} 10 GeV$^2$ the DIS $ep\to eX$ process considered
is dominated by the usual direct $\gamma^*\equiv\gamma(P^2)$ exchange
cross sections and {\underline{not}} `contaminated' anymore by resolved
contributions.  This furthermore explains the trend of the discrepancies
between the data and our as well as other \cite{ref11,ref12} predictions
which can be traced to the fact that the direct photon contributions
were not calculated as if $\gamma^*(P^2)$ was real, as required by our
consistency condition (ii).  Thus the direct photon contribution was
likely underestimated, particularly at the larger value of $x$, $x=0.6$,
resulting in an overestimate of $x\tilde{f}^{\gamma(P^2)}(x,Q^2)$ at
$P^2$ \raisebox{-0.1cm}{$\stackrel{>}{\sim}$} 5 GeV$^2$.
%
%

\section{Summary and Conclusions}
The main purpose of the present paper was to formulate a consistent
set of boundary conditions which allow for a perturbatively stable
LO and NLO calculation of the photonic parton distributions
$f^{\gamma(P^2)}(x,Q^2)$ as well as for a smooth transition to the
parton densities of a real $(P^2=0)$ photon.  Employing the recently
updated \cite{ref21} pionic distributions $f^{\pi}(x,Q^2)$, required
for describing, via VMD, the nonpointlike hadronic components of a 
photon, we arrive at essentially parameter--{\underline{free}}
predictions for $f^{\gamma(P^2)}(x,Q^2)$ which are furthermore in
good agreement with all present measurements of the structure function
$F_2^{\gamma}(x,Q^2)$ of real photons $\gamma\equiv\gamma(P^2=0)$.
It should be noted that the experimentally \mbox{almost} unconstrained pionic
gluon and sea distributions, $g^{\pi}(x,Q^2)$ and $\bar{q}\,^{\pi}(x,Q^2)$,
have been uniquely derived \cite{ref21,ref22} from the experimentally 
rather well known pionic valence density $v^{\pi}(x,Q^2)$ and the 
(also recently updated \cite{ref23} dynamical) parton distributions
of the proton.  We have furthermore implemented these hadronic 
components by using a VMD ansatz for a coherent superposition of 
vector mesons which maximally enhances the contributions of the 
up--quarks to $F_2^{\gamma}$ as favored by all present data.  Since
these hadronic contributions are generated from the valence--like 
input parton distributions at the universal target--mass independent
low resolution scale $Q_0^2=\mu^2\simeq 0.3$ GeV$^2$, we arrive, at
least for real ($P^2=0$) photons, at unique small--$x$ predictions 
for $x$ \raisebox{-0.1cm}{$\stackrel{<}{\sim}$} $10^{-2}$ at $Q^2>\mu^2$
which are of purely dynamical origin, as in the case of hadrons. 
Furthermore, since our universal input scale $\mu^2$ fixes also
uniquely the perturbative pointlike part of the photonic parton
distributions, which dominates for $x$ \raisebox{-0.1cm}{$\stackrel
{>}{\sim}$} 0.1, the large--$x$ behavior of photonic structure functions
is unambiguously predicted as well.

Our expectations for the parton content of virtual ($P^2\neq0$) photons
are clearly more speculative, depending on how one models the hadronic
component (input) of a virtual photon.  The latter is usually assumed
to be similar to the VMD input for a real photon, times a dipole
suppression factor which derives from an effective vector--meson
$P^2$--propagator, cf.\ eq.\ (24).  Whenever a virtual photon is probed
at a scale $Q^2\gg P^2$, with its virtuality being entirely taken 
care of by the (transverse) equivalent photon flux factor, it has to
be considered as a real photon in the sense that cross sections of
subprocesses involving $\gamma(P^2)$ should be calculated as if $P^2=0$.
In other words, the treatment and expressions for $f^{\gamma(P^2)}(x,Q^2)$
as {\underline{on}}--shell transverse partons obeying the usual RG
$Q^2$--evolution equations (with the usual splitting functions of 
{\underline{real}} photons, etc.) dictate an identification of the 
relevant resolved sub--cross--sections $f^{\gamma(P^2)}X\to X'$ with
that of the {\underline{real}} photon, $\hat{\sigma}(f^{\gamma(P^2)}
X\to X')=\hat{\sigma}(f^{\gamma}X\to X')$.  In particular, the calculation
of $F_2^{\gamma(P^2)}(x,Q^2)$ requires the {\underline{same}} photonic
Wilson coefficient $C_{\gamma}(x)$ as for $P^2=0$, in contrast to 
what has been originally proposed \cite{ref7,ref8}.  This allows to
formulate similar boundary conditions in LO and NLO which give rise
to perturbatively stable parton distributions, cross sections (i.e.\
also structure functions) of virtual photons $\gamma(P^2)$ as long
as they are probed at scales $Q^2\gg P^2$ where $Q^2\equiv 4m_c^2,\,\,
\left(p_T^{\rm{jet}}\right)^2$, etc., and typically \mbox{$P^2$ \raisebox
{-0.1cm}{$\stackrel{<}{\sim}$} $\frac{1}{10}\,Q^2$.}  It should be
emphasized that only in this latter kinematic range $P^2\ll Q^2$ is
the whole concept of RG resummed parton distributions of (resolved)
virtual photons appropriate and relevant.  Parton distributions of
virtual photons extracted recently from DIS $ep$ dijet data are in
good agreement with our (parameter--free) predictions.

	Finally, we present simple analytic parametrizations of our
predicted LO and NLO(DIS$_{\gamma}$) parton distributions of real
photons.  From these LO parametrizations one can easily obtain also
the ones for a virtual photon which, whithin sufficient accuracy,
may also be used in NLO.  Our NLO(DIS$_{\gamma}$) parametrizations
of the parton densities of the real photon can be easily transformed
to the $\overline{\rm{MS}}$ scheme according to eq.\ (4) which might
be relevant for future NLO analyses of resolved photon contributions
to hard processes where most NLO subprocesses have so far been
calculated in the $\overline{\rm{MS}}$ scheme.

	A FORTRAN package containing our most recent parametrizations
can be obtained by electronic mail on request.
\vspace{1.5cm}

\noindent{\large\bf{Acknowledgements}}

\noindent This work has been supported in part by the `Bundesministerium 
f\"ur Bildung, Wissenschaft, Forschung und Technologie', Bonn.
\newpage

\noindent{\Large{\bf{Appendix}}}
\renewcommand{\theequation}{A.\arabic{equation}}
\setcounter{equation}{0}

\noindent  Simple analytic parametrizations in LO and NLO of the `
hadronic' piece of the real and virtual photonic parton distributions, 
being proportional to $f^{\pi}(x,Q^2)$ in eqs.\ (13) and (25), respectively,
are already known according to the recently published \cite{ref21}
parametrizations for $f^{\pi}(x,Q^2)$.  Therefore we only need to
parametrize the remaining `pointlike' components in eqs.\ (13) and
(25).

\noindent {\bf{1.}}  Parametrization of LO `pointlike' photonic
parton distributions\\
\vspace{-0.8cm}

\noindent In LO the $Q^2$ dependence of the `pointlike' 
$f^{\gamma}_{pl}(x,Q^2)$
term in (13) enters, apart from an overall $1/\alpha_s(Q^2)$ factor,
merely via the combination 
$L\equiv\alpha_s(Q^2)/\alpha_s(\mu_{\rm{LO}}^2)$ 
as is evident, for example, from eq.\ (2.12) of the first article
in ref.\ \cite{ref1}. Therefore, we prefer to parametrize the
quantity $f_{pl}^{\gamma}(x,Q^2)$ in terms of
\begin{equation}
s\equiv ln\, \frac{ln\,[Q^2/(0.204\, {\rm{GeV}})^2]}
    {ln\,[\mu_{\rm{LO}}^2/(0.204\, {\rm{GeV}})^2]}
\end{equation}

\noindent where \cite{ref23} $\mu_{\rm{LO}}^2=0.26$ GeV$^2$, which
will later provide us a parametrization also for the virtual `pointlike'
component in (25).  Our resulting `pointlike' distributions in eq.\ (13)
can be expressed by the following simple parametrizations, valid for
$0.5$ \raisebox{-0.1cm}{$\stackrel{<}{\sim}$} $Q^2$ 
\raisebox{-0.1cm} {$\stackrel{<}{\sim}$} $10^5$ GeV$^2$ 
(i.e.\  0.31 \raisebox{-0.1cm}{$\stackrel{<}{\sim}$} $s$  
\raisebox{-0.1cm}{$\stackrel{<}{\sim}$} $2.2$) and 
$10^{-5}$ \raisebox{-0.1cm}{$\stackrel{<}{\sim}$} $x < 1$ :   
\begin{eqnarray}
\frac{1}{\alpha}\, x\, f_{pl}^{\gamma}(x,Q^2) & = & \frac{9}{4\pi}\, ln\,
  \frac{Q^2}{(0.204\,\,{\rm{GeV}})^2}\, \Bigg[ s^{\alpha}x^a(A+B\sqrt{x} 
     + C\,x^b)  \nonumber\\ 
 & &    + s^{\alpha'}{\rm{exp}}\Bigg( -E+\sqrt{E's^{\beta}ln\,
      \frac{1}{x}}\, \Bigg)\, \Bigg] \, (1-x)^D 
\end{eqnarray}
where for $f_{pl}^{\gamma}=u_{pl}^{\gamma}=\bar{u}_{pl}^{\gamma}$ 
\begin{equation}
\begin{array}{lcllcl}
\alpha & = & 0.897, & \alpha' & = & 2.626,\\[1mm]
\beta & = & 0.413, & & &\\[1mm]
a & = & 2.137 - 0.310\,\sqrt{s}, &  b & = & -1.049 + 0.113\, s,\\[1mm]
A & = & -0.785 + 0.270\,\sqrt{s}, &  B & = &  0.650 - 0.146\, s,\\[1mm]
C & = & 0.252 - 0.065\,\sqrt{s}, &  D & = & -0.116 + 0.403\, s -
   0.117\,s^2,\\[1mm]
E & = & 6.749 + 2.452\,s - 0.226\, s^2, & E' &  = & 1.994\,s - 0.216\,s^2\,,
\end{array}
\end{equation}
\newpage
for $f_{pl}^{\gamma} = d_{pl}^{\gamma} = \bar{d}\,^{\gamma}_{pl}
  = s_{pl}^{\gamma} = \bar{s}\,^{\gamma}_{pl}$
\begin{equation}
\begin{array}{lcllcl}
\alpha & = & 1.084, & \alpha' & = & 2.811,\\[1mm]
\beta & = & 0.960, & & &\\[1mm]
a & = & 0.914,  &  b & = & 3.723 - 0.968\, s,\\[1mm]
A & = & 0.081 - 0.028\,\sqrt{s}, &  B & = &  -0.048\\[1mm]
C & = & 0.094 - 0.043\,\sqrt{s}, &  D & = &  0.059 + 0.263\, s 
          - 0.085\,s^2,\\[1mm]
E & = & 6.808 + 2.239\,s - 0.108\, s^2, & E' &  = & 1.225 + 0.594\,s
          - 0.073\,s^2\,,
\end{array}
\end{equation}
and for $f_{pl}^{\gamma}=g_{pl}^{\gamma}$
\begin{equation}
\begin{array}{lcllcl}
\alpha & = & 1.262, & \alpha' & = & 2.024,\\[1mm]
\beta & = & 0.770, & & &\\[1mm]
a & = & 0.081,  &  b & = & 0.848\\[1mm]
A & = & 0.012 + 0.039\,\sqrt{s}, &  B & = &  -0.056 - 0.044\,s\\[1mm]
C & = & 0.043 + 0.031\,s, &  D & = &  0.925 + 0.316\,s,\\[1mm]
E & = & 3.129 + 2.434\,s - 0.115\, s^2, & E' &  = & 1.364 + 1.227\,s
          - 0.128\,s^2\,.
\end{array}
\end{equation}

With these parametrizations at hand in terms of $s$ in (A.1), the
appropriate ones for the `pointlike' distributions $f_{pl}^{\gamma(P^2)}
(x,Q^2)$ of a {\underline{virtual}} photon appearing in eq.\ (25) are 
simply given by the same expressions above where in (A.1) $\mu_{\rm{LO}}^2$
has to be replaced by $\tilde{P}^2={\rm{max}}(P^2,\mu_{\rm{LO}}^2)$.
As discussed in Sect. 3, these parametrizations can, within sufficient
accuracy, also be used for the parton distribution of {\underline{virtual}}
photons in NLO.
\vspace{0.5cm}

\noindent {\bf{2.}}  Parametrization of NLO `pointlike' photonic
parton distributions\\
\vspace{-0.8cm}

\noindent In NLO the $Q^2$ dependence of the `pointlike' distributions
of real photons in (13), $f_{pl}^{\gamma}(x,Q^2)$, can be easily
described in terms of the following `effective' logarithmic ratio
\begin{equation}
s\equiv ln\, \frac{ln\,[Q^2/(0.299\, {\rm{GeV}})^2]}
    {ln\,[\mu_{\rm{NLO}}^2/(0.299\, {\rm{GeV}})^2]}
\end{equation}
to be evaluated for $\mu_{\rm{NLO}}^2=0.40$ GeV$^2$.  Our 
NLO(DIS$_{\gamma}$) predictions can now be parametrized as the LO ones
and are similarly valid for 
$0.5$ \raisebox{-0.1cm}{$\stackrel{<}{\sim}$} $Q^2$ \raisebox{-0.1cm}
{$\stackrel{<}{\sim}$} $10^5$ GeV$^2$ (i.e.\ 0.14 \raisebox{-0.1cm}
{$\stackrel{<}{\sim}$} $s$ \raisebox{-0.1cm}{$\stackrel{<}{\sim}$}
$2.38$) and $10^{-5}$ \raisebox{-0.1cm}{$\stackrel{<}{\sim}$} $x < 1$.
For convenience we include now the NLO $\alpha_s$ in the r.h.s.\ of
(A.2), i.e.
\newpage
\begin{eqnarray}
\frac{1}{\alpha}\, x\, f_{pl}^{\gamma}(x,Q^2) & = & \Bigg[ s^{\alpha} x^a
  (A+B\sqrt{x}+Cx^b) \nonumber\\ 
& &  + s^{\alpha'}{\rm{exp}} \Bigg( -E + \sqrt{E's^{\beta}
   \,ln\,\frac{1}{x}}\, \Bigg) \Bigg]\, (1-x)^D
\end{eqnarray}
where for $f_{pl}^{\gamma}=u_{pl}^{\gamma}=\bar{u}_{pl}^{\gamma}$ 
\begin{equation}
\begin{array}{lcllcl}
\alpha & = & 1.051,  & \alpha' & = & 2.107,\\[1mm]
\beta & = & 0.970, & & &\\[1mm]
a & = & 0.412 - 0.115\,\sqrt{s}, &  b & = & 4.544 - 0.563\, s,\\[1mm]
A & = & -0.028\,\sqrt{s} + 0.019\,s^2, &  B & = &  0.263 + 0.137\, s,\\[1mm]
C & = & 6.726 - 3.264\,\sqrt{s} - 0.166\,s^2, & D & = & 1.145 - 0.131\, 
   s^2,\quad\quad\quad\,\,\\ [1mm]
E & = & 4.122 + 3.170\,s - 0.598\, s^2, & E' &  = & 1.615\,s - 0.321\,s^2\,,
\end{array}
\end{equation}
for $f_{pl}^{\gamma} = d_{pl}^{\gamma} = \bar{d}\,^{\gamma}_{pl}
  = s_{pl}^{\gamma} = \bar{s}\,^{\gamma}_{pl}$
\begin{equation}
\begin{array}{lcllcl}
\alpha & = & 1.043, & \alpha' & = & 1.812,\\[1mm]
\beta & = & 0.457, & & &\\[1mm]
a & = & 0.416 - 0.173\,\sqrt{s}, &  b & = & 4.489 - 0.827\, s,\\[1mm]
A & = & -0.010\,\sqrt{s} + 0.006\,s^2, &  B & = &  0.064 + 0.020\,s\\[1mm]
C & = & 1.577 - 0.916\,\sqrt{s}, &  D & = &  1.122 - 0.093\,s - 0.100
         \,s^2,\\[1mm]
E & = & 5.240 + 1.666\,s - 0.234\,s^2, & E' &  = & 1.284\,s
\end{array}
\end{equation}
and for $f_{pl}^{\gamma}=g_{pl}^{\gamma}$
\begin{equation}
\begin{array}{lcllcl}
\alpha & = & 0.901, & \alpha' & = & 1.773,\\[1mm]
\beta & = & 1.666, & & &\\[1mm]
a & = & 0.844 - 0.820\,\sqrt{s}  &  b & = & 2.302 - 0.474\,s\\[1mm]
A & = & 0.194 &  B & = &  -0.324 + 0.143\,s\\[1mm]
C & = & 0.330 - 0.177\,s, &  D & = &  0.778 + 0.502\,s -
         0.154\,s^2\\[1mm]
E & = & 2.895 + 1.823\,s - 0.441\, s^2, & E' &  = & 2.344 - 0.584\,s\, .
\end{array}
\end{equation}

\newpage

\noindent{\Large{\bf{\underline{Figure Captions}}}}
\begin{itemize}
\item[\bf{Fig.\ 1}]  Comparison of our radiatively generated LO and 
	NLO(DIS$_{\gamma}$) predictions for $F_2^{\gamma}(x,Q^2)$, based 
	on the valence--like parameter--free VMD input in eq.\ (9), with
	the data of ref.\ \cite{ref33}.  For our comparison the 
	GRV$_{\gamma}$ \cite{ref1} results are shown as well.  In both 
	cases, the charm contribution has been added, in the relevant 
	kinematic region $W\geq 2m_c$, according to eqs.\ (15) and (16).  

\item[\bf{Fig.\ 2}]  Comparison of our predicted LO and NLO(DIS$_{\gamma}$)
	distributions $u^{\gamma}=\bar{u}\,^{\gamma},\,\, d^{\gamma}=
	\bar{d}\,^{\gamma}$ and $g^{\gamma}$ at $Q^2=10$ GeV$^2$ with
	the LO/NLO GRV$_{\gamma}$ densities \cite{ref1}, the LO SaS 1D
	and 2D \cite{ref11} and the NLO AFG \cite{ref3} distributions.
	The `hadronic' (pionic) components of our total LO and NLO
	results in eq.\ (11) are displayed by the dashed curves. 
	
\item[\bf{Fig.\ 3}]  Detailed small--$x$ (as well as large--$x$) 
	behavior and predictions of our radiatively generated 
	$u^{\gamma}=\bar{u}\,^{\gamma}$ and $g^{\gamma}$ distributions 
	in LO and NLO(DIS$_{\gamma}$) at fixed values of $Q^2$.  The 
	dashed--dotted curves show the hadronic NLO contribution 
	$f_{\rm{had}}^{\gamma}$ to $f^{\gamma}=f_{\rm{pl}}^{\gamma} +
	f_{\rm{had}}^{\gamma}$ in eq.\ (11).  The valence--like inputs
	at $Q^2=\mu_{\rm{LO,\,NLO}}^2$, according to eq.\ (9), are
	shown by the lowest curves referring to $\mu^2$.  For comparison
	we show the steeper NLO GRV$_{\gamma}$ \cite{ref1} expectations
	as well.  The results have been multiplied by the number 
	indicated in brackets.

\item[\bf{Fig.\ 4}] Comparison of our LO and NLO predictions for
	$xg^{\gamma}$ at $Q^2\equiv \langle (p_T^{\rm{jet}})^2 \rangle
	= 75$ GeV$^2$ with HERA(H1) measurements \cite{ref34}.  The
	GRV$_{\gamma}$ and SaS expectations are taken from refs.\ 
	\cite{ref1} and \cite{ref11}, respectively.

\item[\bf{Fig.\ 5}] LO and NLO predictions for $F_{\rm{eff}}^{\gamma(P^2)}
	\equiv F_2^{\gamma(P^2)} + \frac{3}{2}F_L^{\gamma(P^2)}$.  The
	charm contributions have been calculated according to eqs.\ (15),
	(16) and (28), and the `resolved' charm contribution to 
	$F_L^{\gamma(P^2)}$ is calculated analogously as in (16) with
	$f_L^{\gamma^*(Q^2)g^{\gamma}\to h\bar{h}}(x,Q^2)$ being given
	by eq.\ (28) with $e_c^4\alpha\to e_c^2\alpha_s(\mu_F^2)/6$.
	The `box' is defined by eqs.\ (26) and (27).  The PLUTO data 
	are taken from ref.\ \cite{ref39}.
\newpage
    
\item[\bf{Fig.\ 6a}] LO and NLO predictions for the $x$ dependence of the
	virtual photon structure function $F_2^{\gamma(P^2)}$ at 
	$Q^2=10$ GeV$^2$ and various fixed values of $P^2\ll Q^2$,
	neglecting any heavy quark contribution.  For comparison we also
	show the NLO GRS \cite{ref10} results as well as the predictions
	for a real ($P^2=0$) photon.  Notice that the dotted curve for
	$P^2=0$ referred to as GRS obviously coincides with the 
	GRV$_{\gamma}$ result \cite{ref1}.  The results have been
	multiplied by the number indicated in brackets.

\item[\bf{Fig.\ 6b}] As in fig.\ 6a but at $Q^2=100$ GeV$^2$.

\item[\bf{Fig.\ 7a}] Comparison of our predicted LO and NLO(DIS$_{\gamma})$
	distributions of the virtual photon at $Q^2=10$ GeV$^2$ and various
	fixed values of $P^2\ll Q^2$ with the GRS densities \cite{ref10}.
	The curves refer from top to bottom to $P^2=0,\,\, 0.2$ and 1 
	GeV$^2$, respectively.  The results for the real photon ($P^2=0$)
	are very similar to the ones in fig.\ 2 where the GRV$_{\gamma}$
	curves practically coincide with GRS.

\item[\bf{Fig.\ 7b}] Our LO distributions as in fig.\ 7a compared with
	the ones of SaS \cite{ref11}.

\item[\bf{Fig.\ 8}] LO and NLO predictions for the up--quark and 
	gluon distributions
	of a virtual photon $\gamma(P^2)$ at $Q^2=10$ GeV$^2$.  For 
	comparison the results for the real photon ($P^2=0$) are shown as
	well.  The NLO `hadronic' contribution in (25) is also shown
	separately.  The GRS expectations are taken from ref.\ \cite{ref10}.
	The DIS$_{\gamma}$ results for $u^{\gamma(P^2)}$ can be easily
	converted to the $\overline{\rm{MS}}$ scheme with the help of
	eq.\ (4), whereas $g^{\gamma(P^2)}$ remains the same in both
	schemes.  The results have been multiplied by the numbers indicated
	in brackets.

\item[\bf{Fig.\ 9}] As in fig.\ 8 but at $Q^2=100$ GeV$^2$.

\item[\bf{Fig.\ 10}] Predictions for the LO effective parton density 
	$x\tilde{f}\,^{\gamma(P^2)}(x,Q^2)$, defined in eq.\ (29), at
	the scale $Q^2\equiv\left(p_T^{\rm{jet}}\right)^2=85$ GeV$^2$
	and at two fixed values of $x$.  The H1 data \cite{ref14} have
	been extracted from DIS ep dijet production.  The solid curves
	refer to our predictions in the theoretically legitimate region
	$P^2\ll Q^2\equiv\left( p_T^{\rm{jet}}\right)^2$, whereas the
	dashed curves extend into the kinematic region of larger $P^2$
	approaching $Q^2$ where the concept of parton distributions of
	virtual photons is not valid anymore (see text).
\end{itemize}

\newpage
\pagestyle{empty}
\begin{figure}
\centering
\vspace*{-1cm}
\epsfig{figure=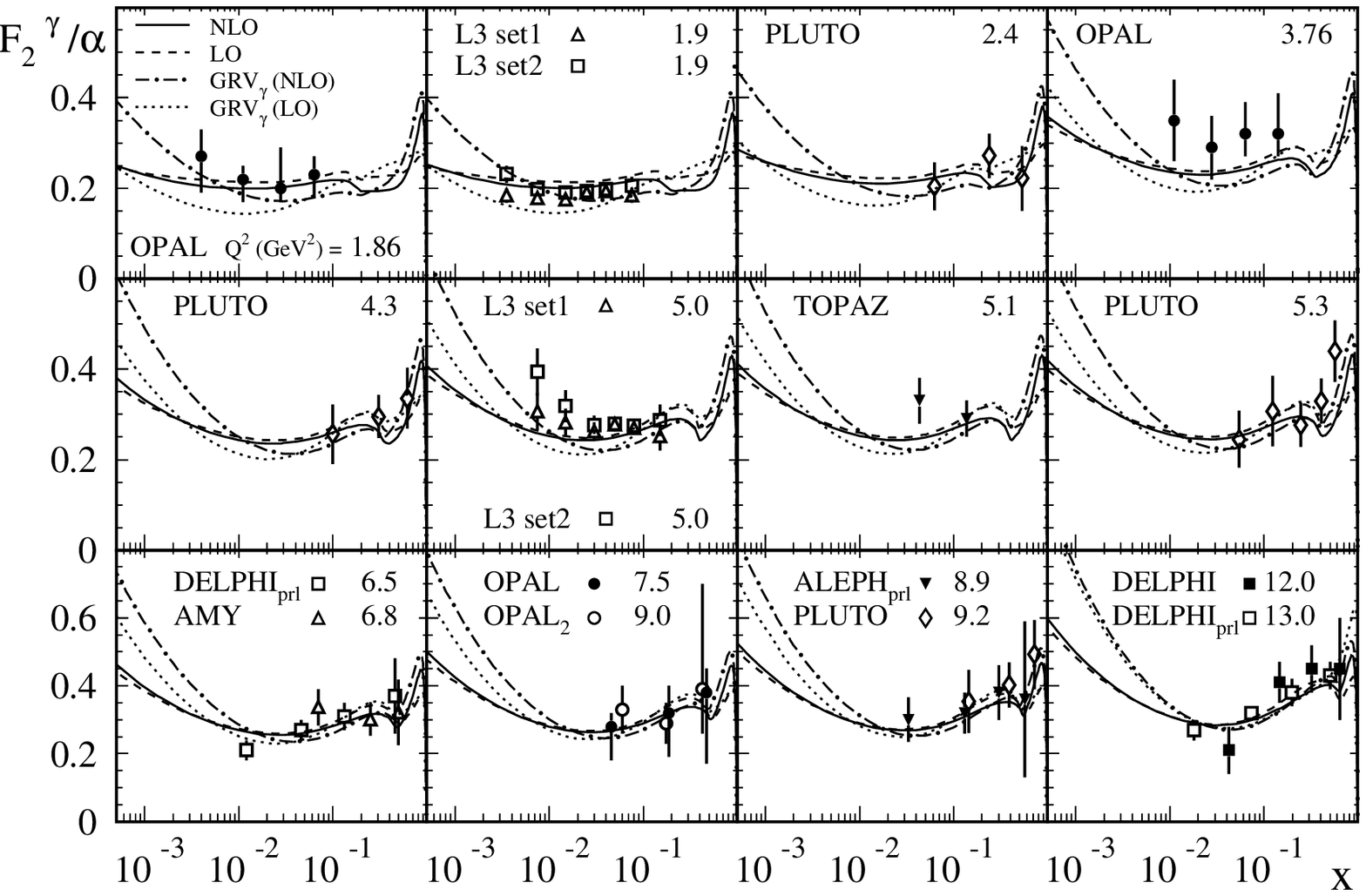,width=16cm}

\end{figure}
\begin{figure}
\centering
\vspace*{-0.5cm}
\epsfig{figure=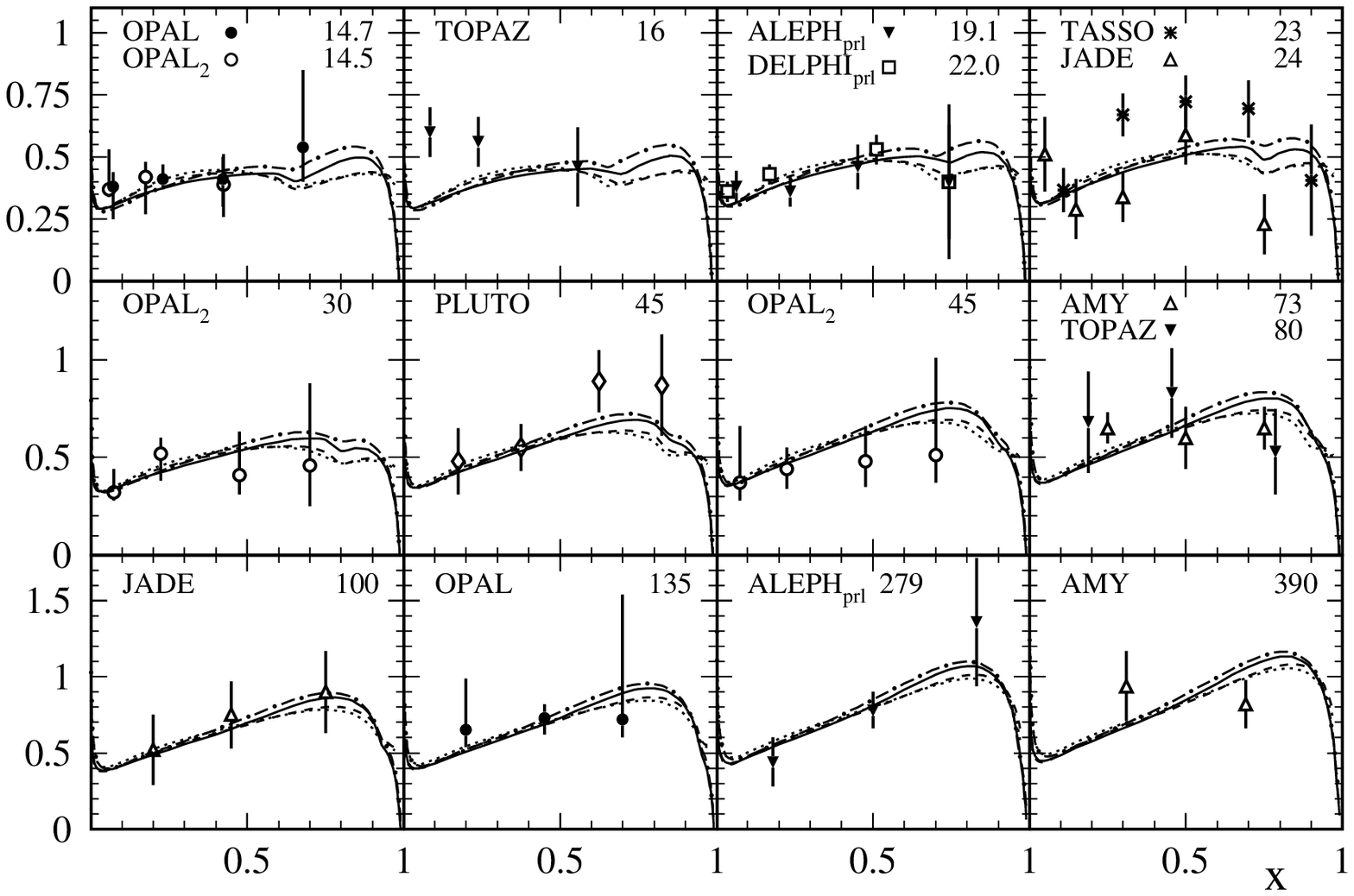,width=16cm}

\vspace*{0.3cm}
{\large\bf Fig. 1}
\end{figure}

\newpage
\begin{figure}[t]
\centering
\epsfig{figure=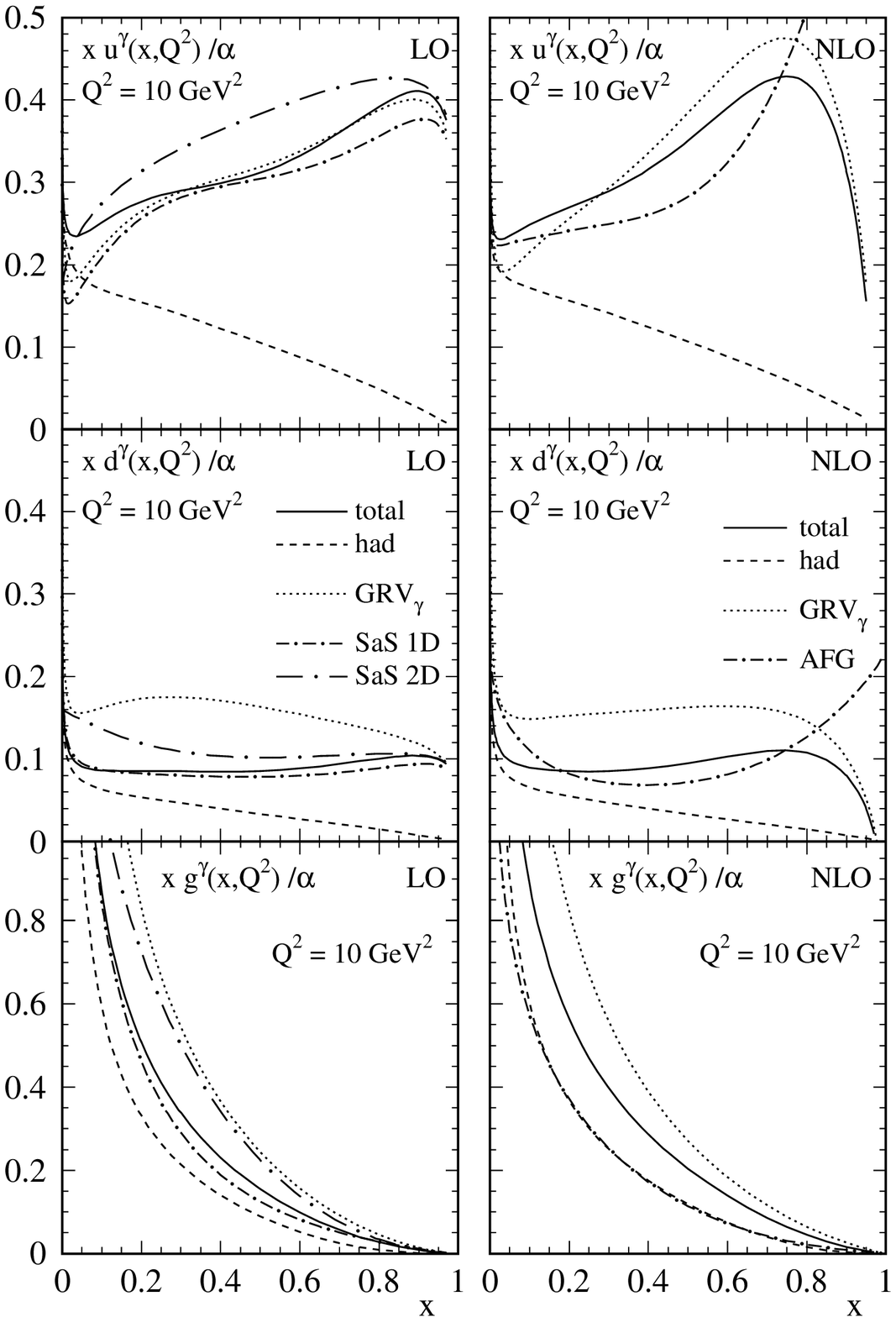,width=12cm}

\vspace*{1cm}
{\large\bf Fig. 2}
\end{figure}

\newpage
\begin{figure}[t]
\centering
\epsfig{figure=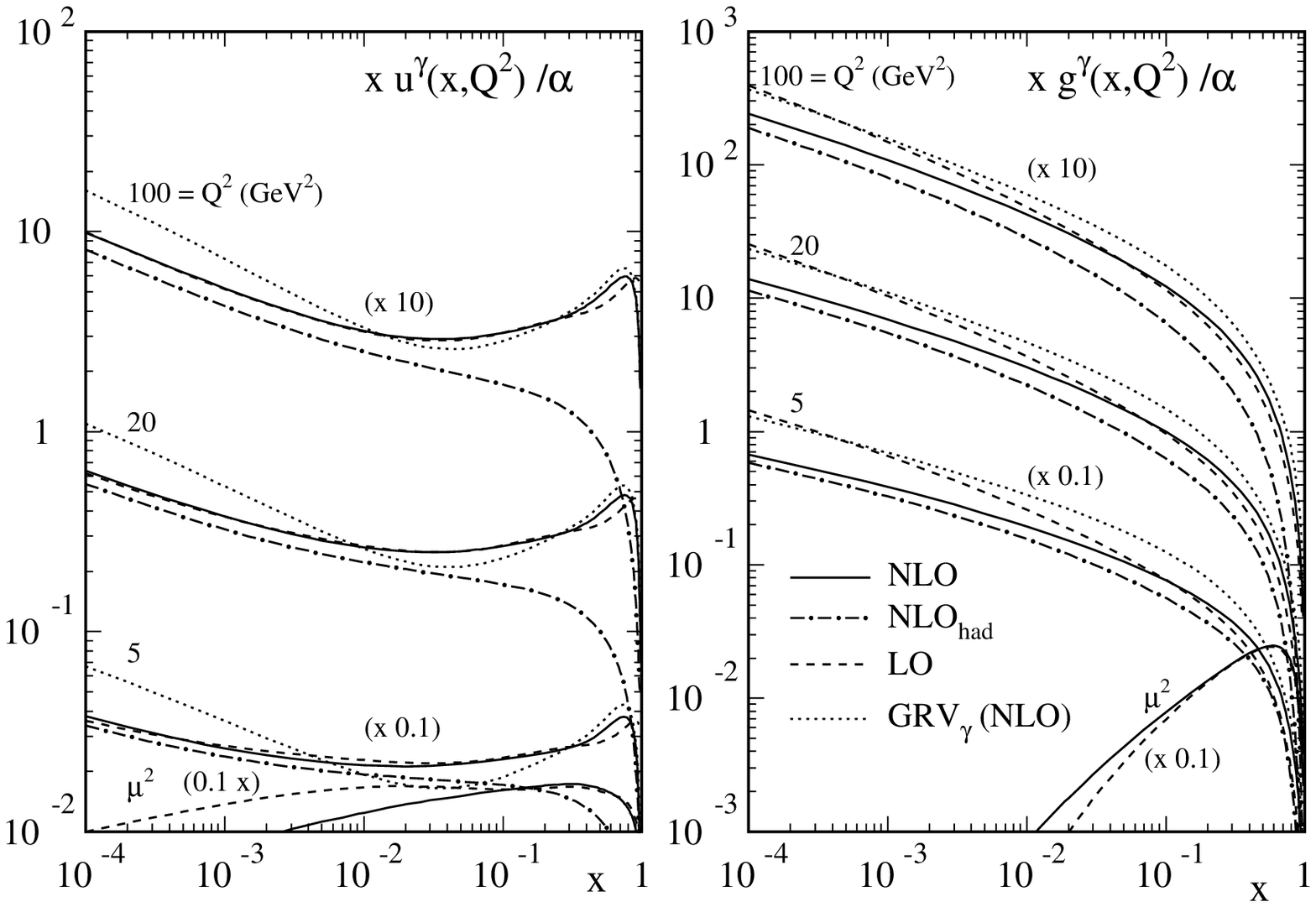,angle=90,width=12cm}

\vspace*{1cm}
{\large\bf Fig. 3}
\end{figure}

\newpage
\begin{figure}[t]
\centering
\epsfig{figure=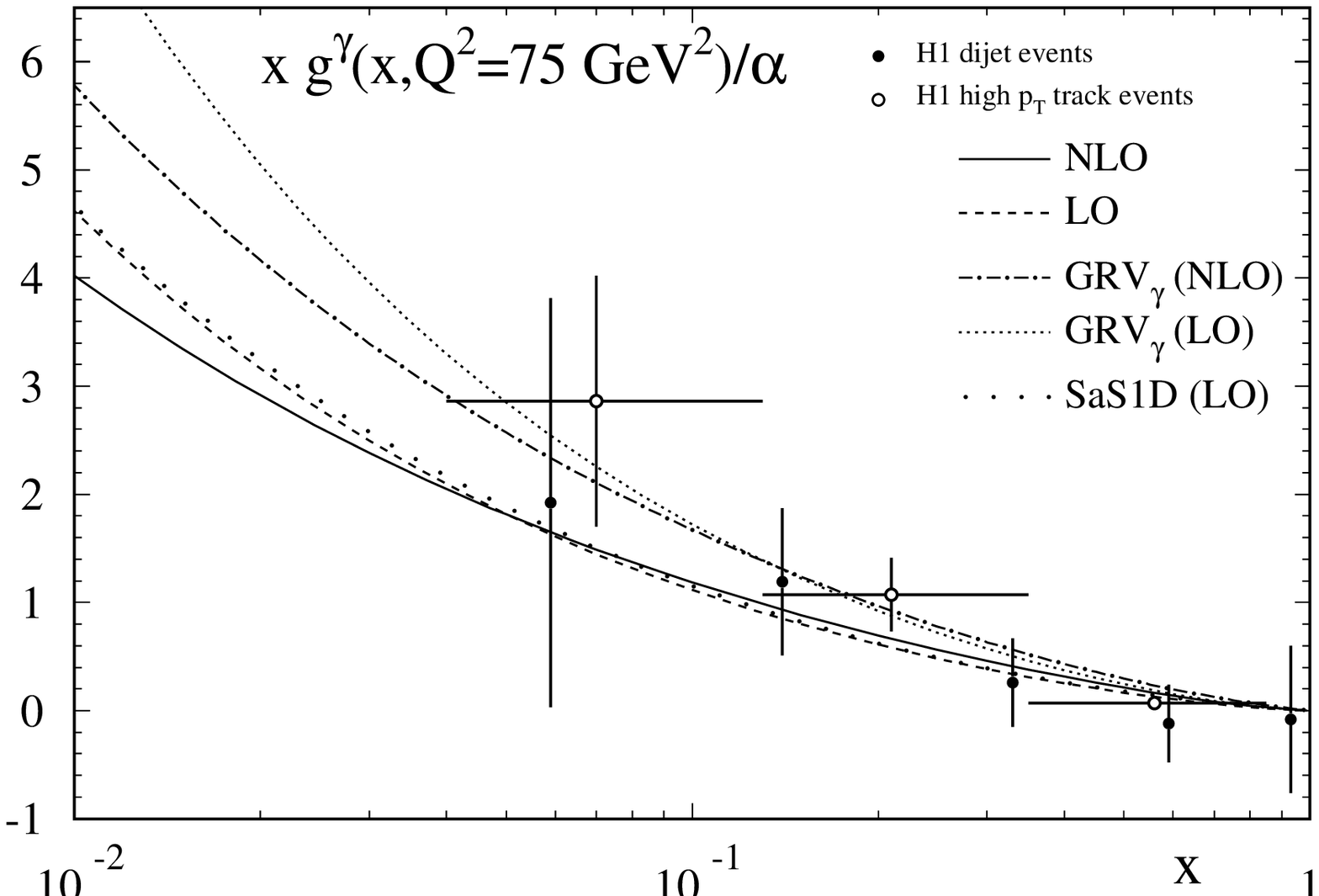,width=14cm}

\vspace*{0.5cm}
{\large\bf Fig. 4}
\end{figure}

\begin{figure}[t]
\centering
\epsfig{figure=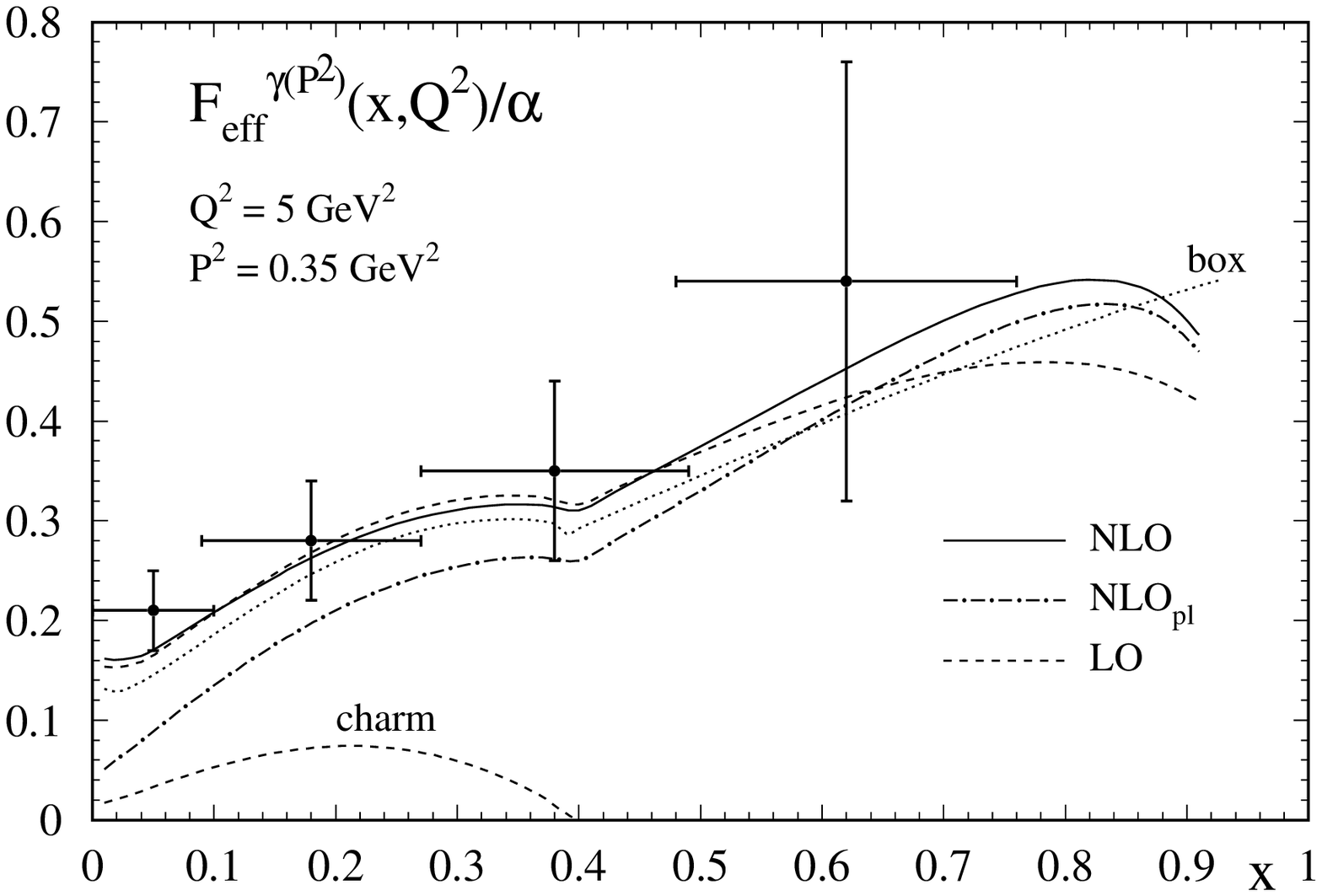,width=14cm}

\vspace*{0.5cm}
{\large\bf Fig. 5}
\end{figure}

\newpage
\begin{figure}[t]
\centering
\epsfig{figure=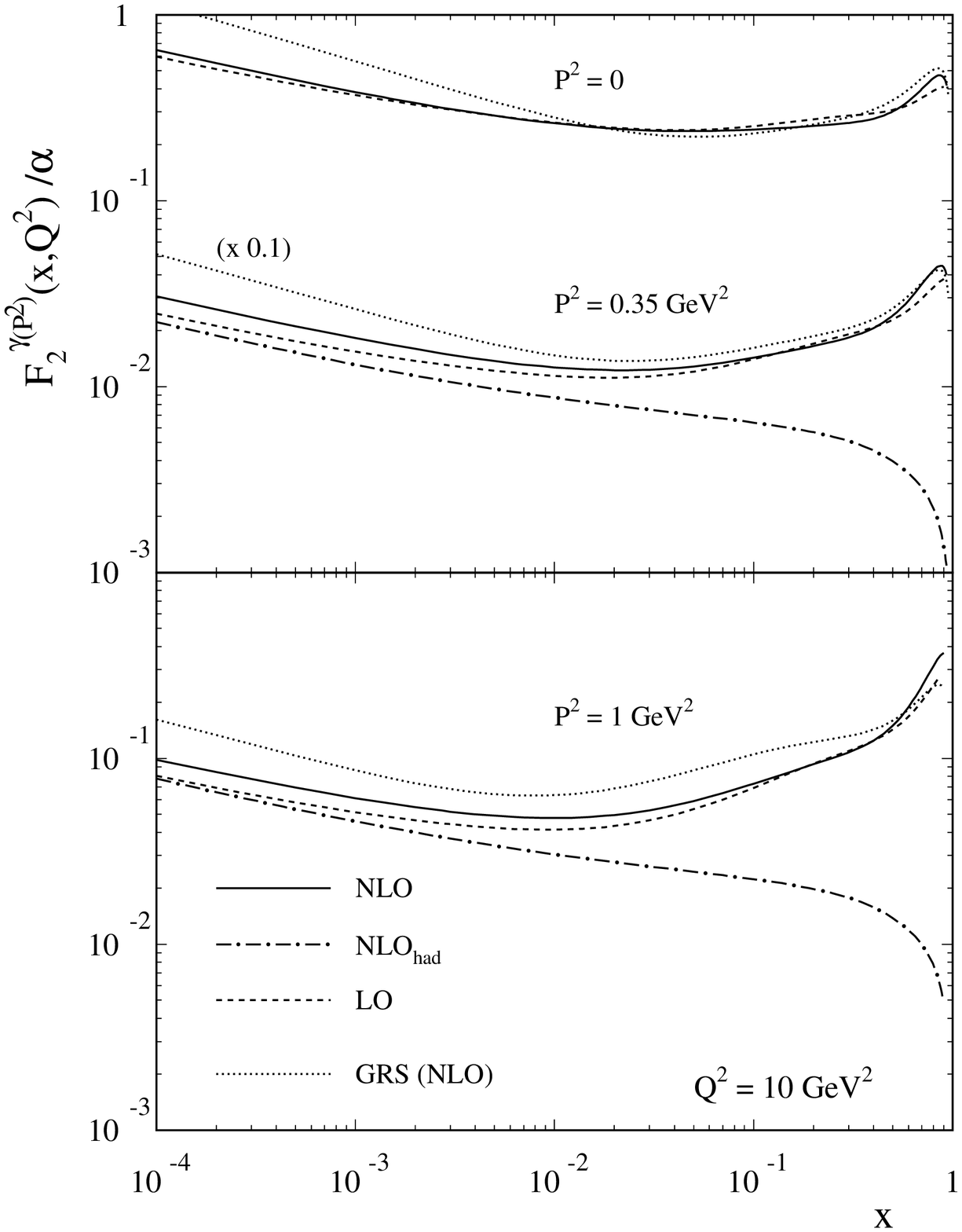,width=16cm}

\vspace*{-0.5cm}
{\large\bf Fig. 6a}
\end{figure}

\newpage
\begin{figure}[t]
\centering
\epsfig{figure=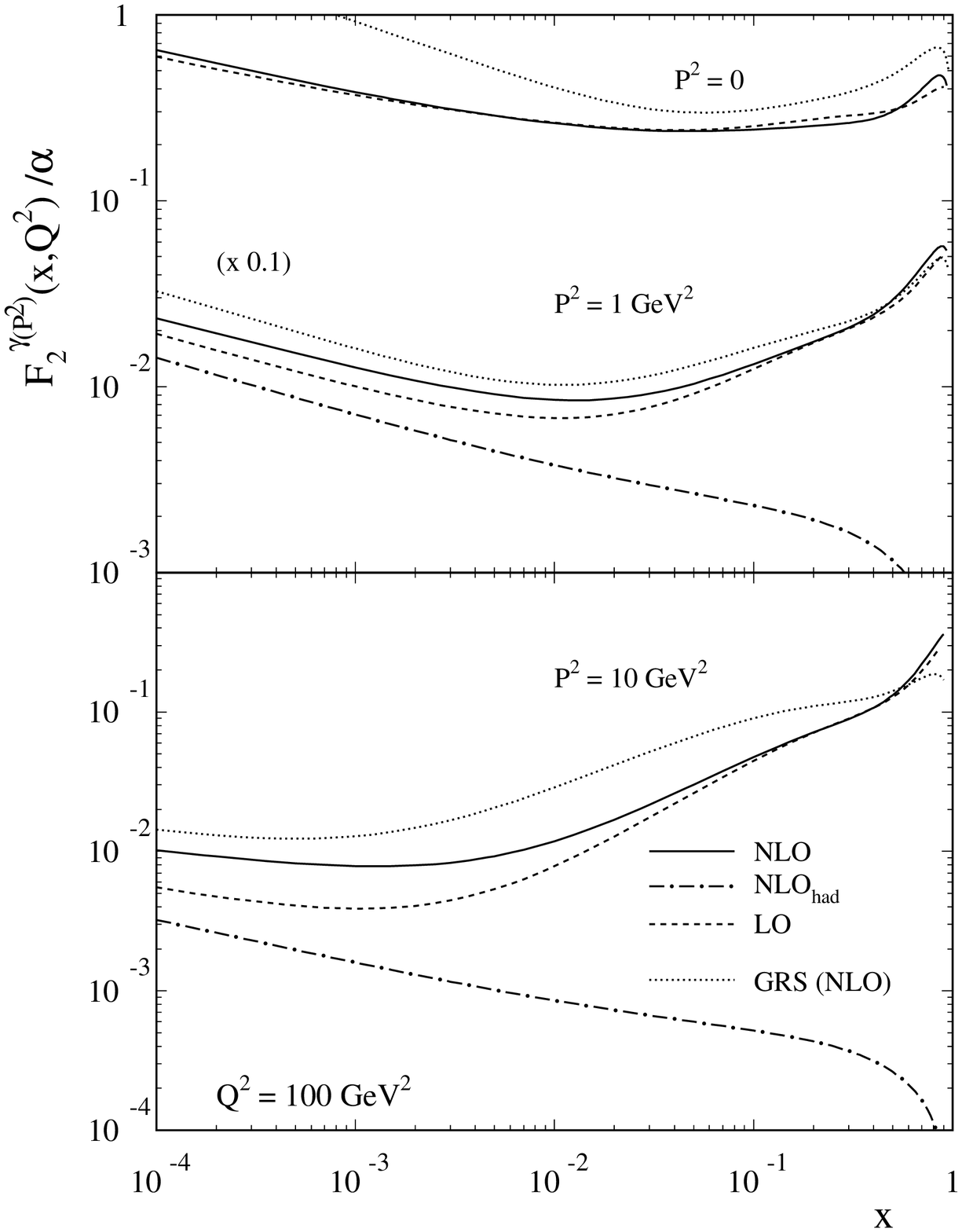,width=16cm}

\vspace*{-0.5cm}
{\large\bf Fig. 6b}
\end{figure}

\newpage
\begin{figure}[t]
\centering
\epsfig{figure=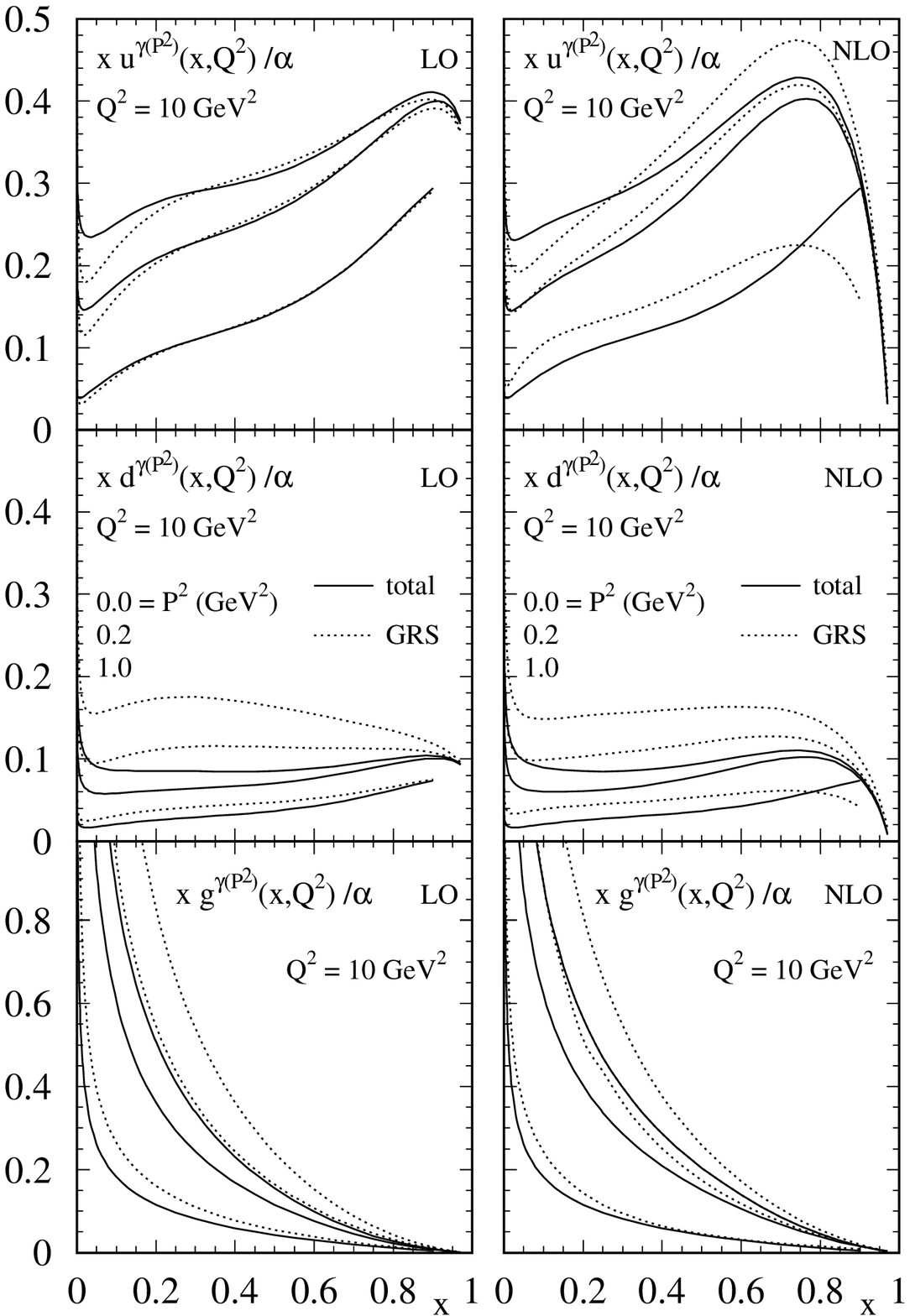,width=12cm}

\vspace*{1cm}
{\large\bf Fig. 7a}
\end{figure}

\newpage
\begin{figure}[t]
\centering
\epsfig{figure=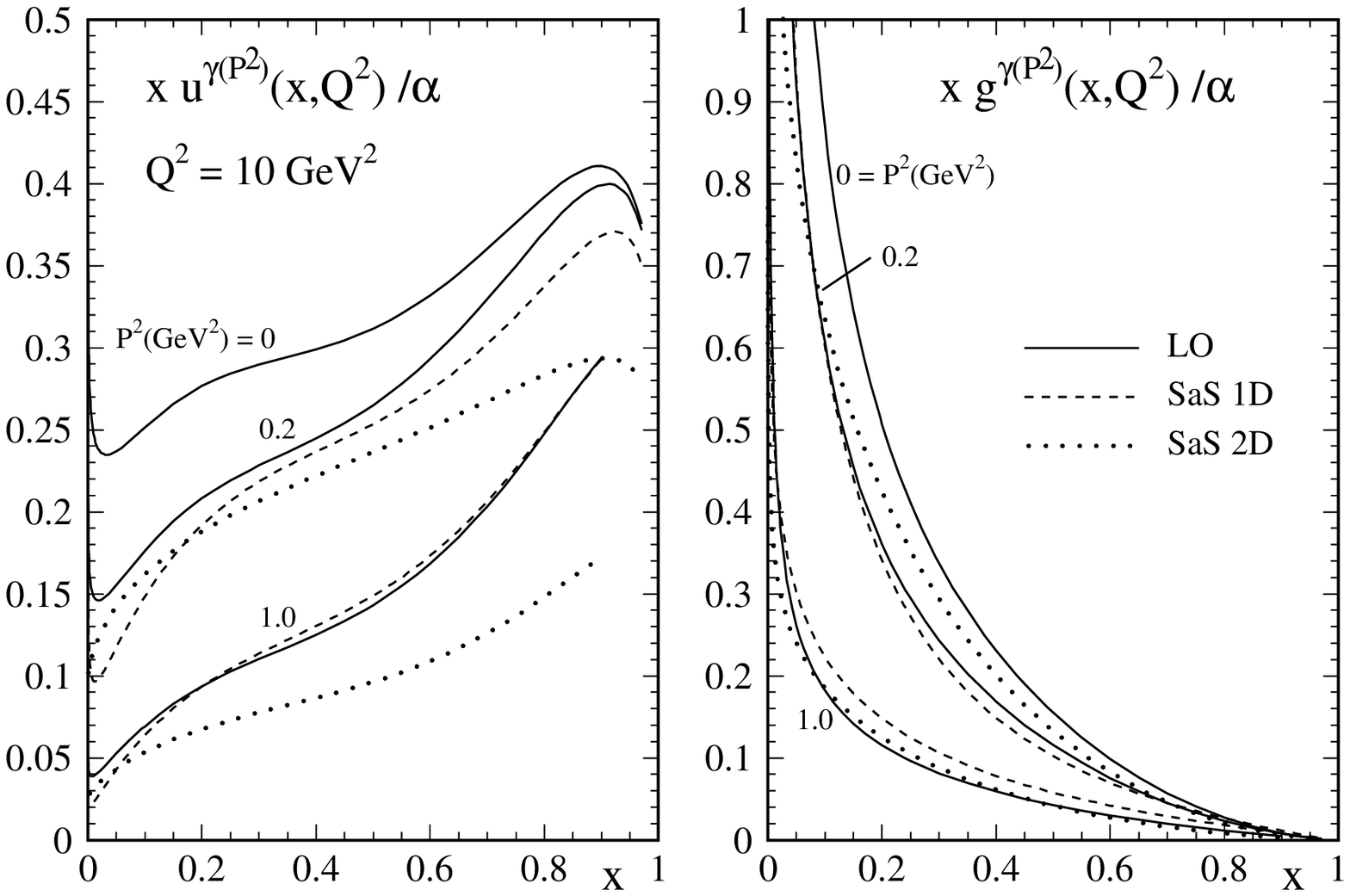,angle=90,width=12cm}

\vspace*{1cm}
{\large\bf Fig. 7b}
\end{figure}

\newpage
\begin{figure}[t]
\centering
\epsfig{figure=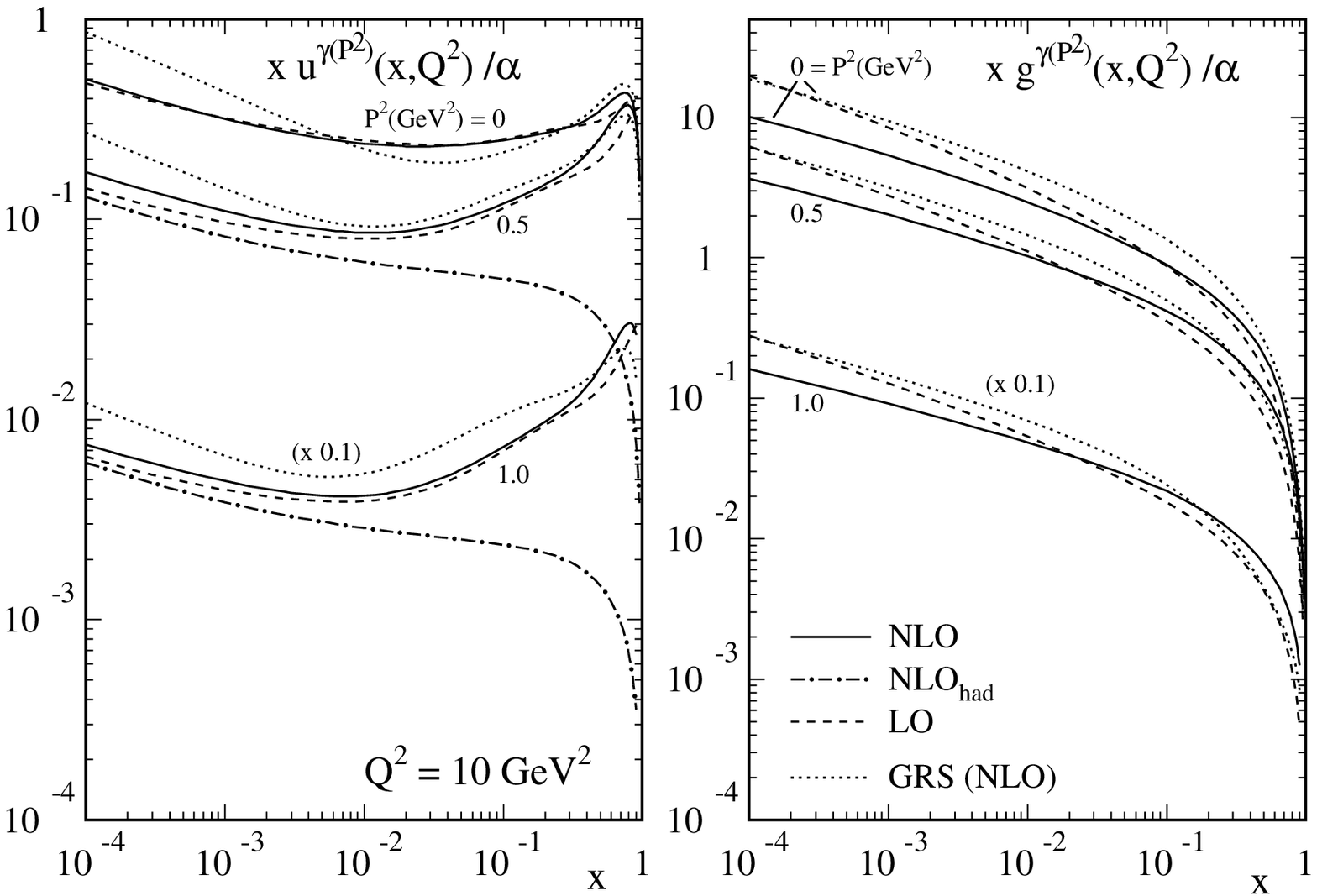,angle=90,width=12cm}

\vspace*{1cm}
{\large\bf Fig. 8}
\end{figure}

\newpage
\begin{figure}[t]
\centering
\epsfig{figure=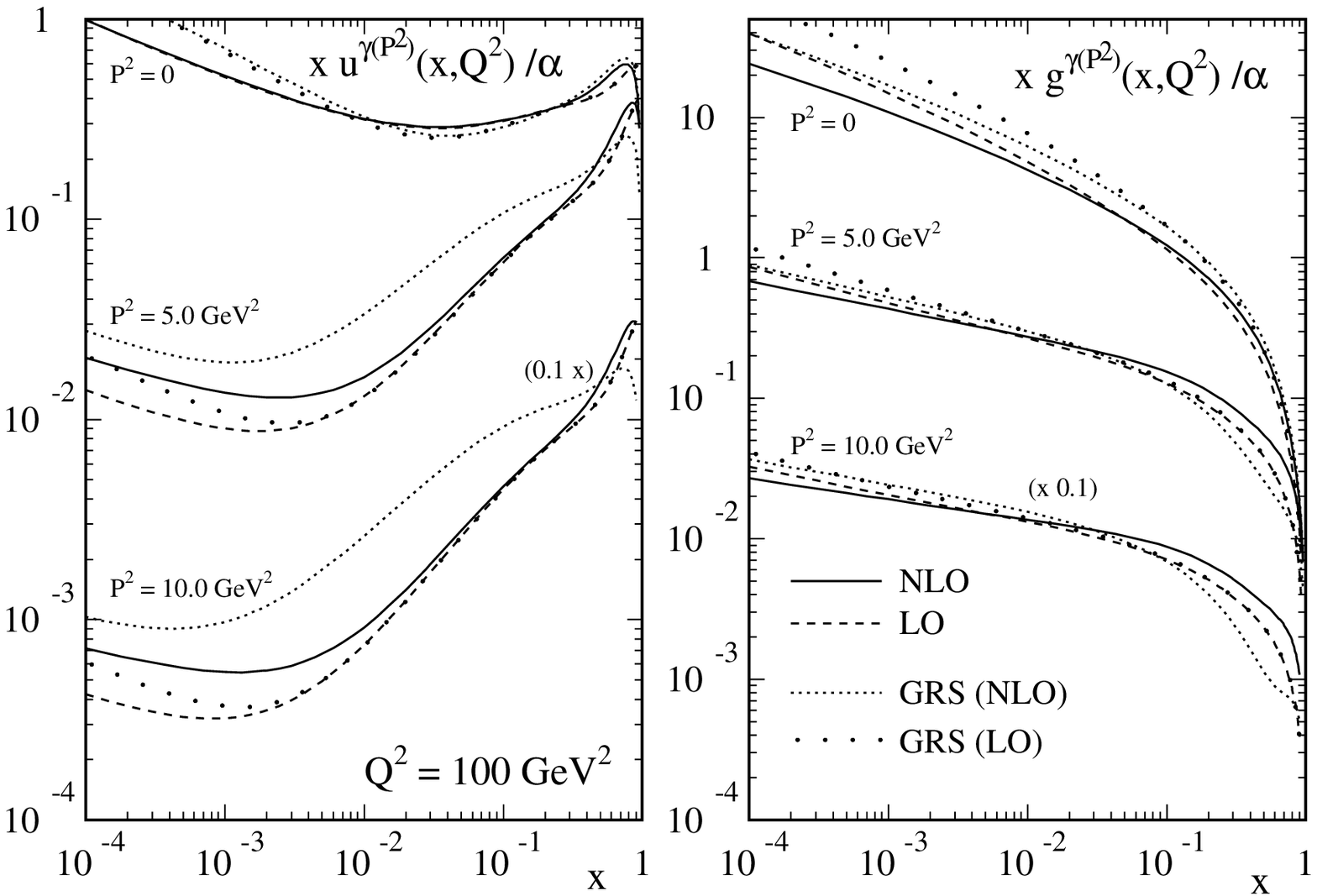,angle=90,width=12cm}

\vspace*{1cm}
{\large\bf Fig. 9}
\end{figure}

\newpage
\begin{figure}[t]
\centering
\epsfig{figure=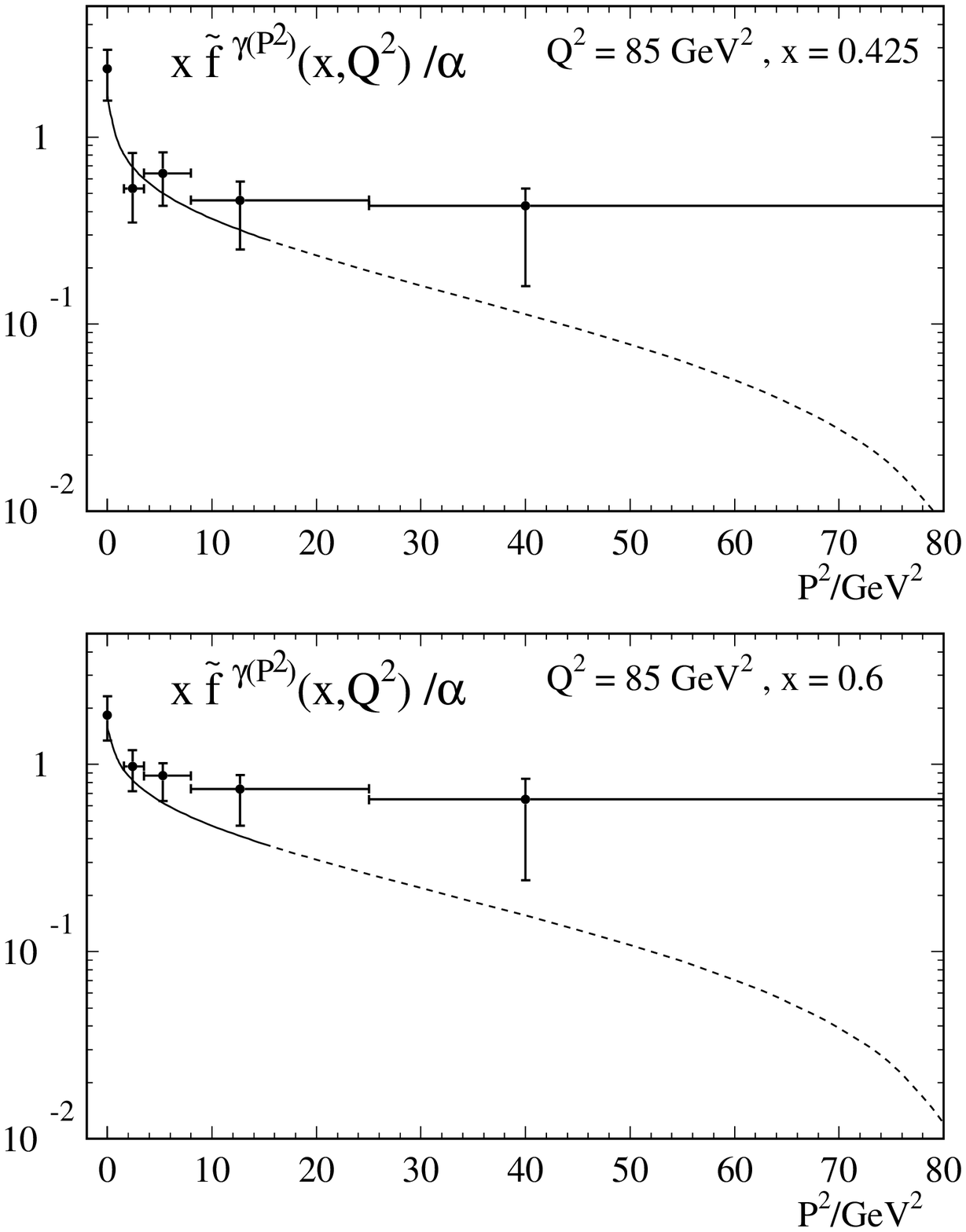,width=15cm}

\vspace*{1cm}
{\large\bf Fig. 10}
\end{figure}
\end{document}